\documentclass[fleqn,usenatbib]{mnras}

\usepackage{amsfonts}

\usepackage[T1]{fontenc}
\usepackage{ae,aecompl}
\usepackage[position=top]{subfig}
\usepackage{float}
\usepackage{comment}

\usepackage{grffile}
\usepackage{etoolbox}
\makeatletter 
  \patchcmd{\NAT@citex}
    {\@citea\NAT@hyper@{%
      \NAT@nmfmt{\NAT@nm}%
      \hyper@natlinkbreak{\NAT@aysep\NAT@spacechar}{\@citeb\@extra@b@citeb}%
      \NAT@date}}
    {\@citea\NAT@nmfmt{\NAT@nm}%
    \NAT@aysep\NAT@spacechar\NAT@hyper@{\NAT@date}}{}{}

  \patchcmd{\NAT@citex}
    {\@citea\NAT@hyper@{%
      \NAT@nmfmt{\NAT@nm}%
      \hyper@natlinkbreak{\NAT@spacechar\NAT@@open\if*#1*\else#1\NAT@spacechar\fi}%
        {\@citeb\@extra@b@citeb}%
      \NAT@date}}
    {\@citea\NAT@nmfmt{\NAT@nm}%
    \NAT@spacechar\NAT@@open\if*#1*\else#1\NAT@spacechar\fi\NAT@hyper@{\NAT@date}}
    {}{}
\makeatother

\usepackage{etoolbox}
 \makeatother
\usepackage{graphicx}	
\usepackage{amsmath}	
\usepackage{amssymb}	
\usepackage{pifont}
\newcommand{\FeII}{Fe\ \textsc{ii}}
\newcommand{\HeII}{He\ \textsc{ii}}
\newcommand{\SiII}{Si\ \textsc{ii}}
\newcommand{\CII}{C\ \textsc{ii}}
\newcommand{\OI}{O\ \textsc{i}}
\title[Termination of Pop~III star formation]{When did Population~III star formation end?}

\author[B. Liu, V. Bromm]{Boyuan Liu\textsuperscript{\href{https://orcid.org/0000-0002-4966-7450}{\includegraphics[width=2.5mm]{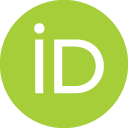}}\,}\thanks{E-mail: boyuan@utexas.edu}$^{1}$ 
and Volker Bromm$^{1}$
\\
$^{1}$Department of Astronomy, University of Texas, Austin, TX 78712, USA\\
}

\date{Accepted XXX. Received YYY; in original form ZZZ}

\pubyear{2020}

\begin{document}
\label{firstpage}
\pagerange{\pageref{firstpage}--\pageref{lastpage}}
\maketitle

\begin{abstract}
We construct a theoretical framework to study Population~III (Pop~III) star formation in the post-reionization epoch ($z\lesssim 6$) by combining cosmological simulation data with semi-analytical models. We find that due to radiative feedback (i.e. Lyman-Werner and ionizing radiation) massive haloes ($M_{\rm halo}\gtrsim 10^{9}\ \rm M_{\odot}$) are the major ($\gtrsim 90$\%) hosts for potential Pop~III star formation at $z\lesssim 6$, where dense pockets of metal-poor gas may survive to form Pop~III stars, under inefficient mixing of metals released by supernovae. Metal mixing is the key process that determines not only when Pop~III star formation ends, but also the total mass, $M_{\rm PopIII}$, of \textit{active} Pop III stars per host halo, which is a crucial parameter for direct detection and identification of Pop~III hosts. Both aspects are still uncertain due to our limited knowledge of metal mixing during structure formation. Current predictions range from early termination at the end of reionization ($z\sim 5$) to continuous Pop~III star formation extended to $z=0$ at a non-negligible rate $\sim 10^{-7}\ \rm M_{\odot}\ yr^{-1}\ Mpc^{-3}$, with $M_{\rm PopIII}\sim 10^{3}-10^{6}\ \rm M_{\odot}$. This leads to a broad range of redshift limits for direct detection of Pop~III hosts, $z_{\rm PopIII}\sim 0.5-12.5$, with detection rates $\lesssim 0.1-20\ \rm arcmin^{-2}$, for current and future space telescopes (e.g. HST, WFIRST and JWST). Our model also predicts that the majority ($\gtrsim 90$\%) of the cosmic volume is occupied by metal-free gas. Measuring the volume filling fractions of this metal-free phase can constrain metal mixing parameters and Pop~III star formation.  
\end{abstract}
\begin{keywords}
early universe -- dark ages, reionization, first stars --  galaxies: dwarf
\end{keywords}



\section{Introduction}
\label{s1}

The `standard model' of early star formation predicts that the first generation of stars, the so-called Population~III (Pop~III), started to form at redshift $z\sim 20-30$ in minihaloes of $M_{\rm halo}\sim 10^{6}\ \rm M_{\odot}$ \citep{abel2002formation, bromm2002, bromm2013}, driven by cooling from $\rm H_{2}$ and $\rm HD$ molecules in primordial gas \citep{johnson2006,lithium}. Although the properties of Pop~III stars are still uncertain in the absence of direct observations, current theoretical models (e.g. \citealt{stacy2013constraining,susa2014mass,hirano2015primordial,machida2015accretion,stacy2016building,hirano2017formation,hosokawa2020}), and indirect observational constraints (e.g. \citealt{frebel2015near,ji2015preserving,hartwig2015,magg2019observational,ishigaki2018initial,yuta2020}) converge on the picture that Pop~III stars are characterized by a top-heavy initial mass function (IMF), covering a few to a few hundred $\rm M_{\odot}$. As a result, they have distinct features compared with present-day stars formed in metal-enriched enrionments, such as bluer spectra with narrow \HeII\ emission lines and higher efficiencies of producing supernovae (SNe) and (binary) black holes (e.g. \citealt{nagao2008photometric,whalen2013finding,kinugawa2014possible,sobral2015evidence,belczynski2017likelihood,ishigaki2018initial}).

A fundamental question regarding Pop~III stars is when in cosmic history this special mode of star formation was terminated, which is intricately linked to our understanding of the feedback mechanisms that not only regulate Pop~III star formation, but also drive cosmic thermal and chemical evolution. An important goal is to provide guidance to observational campaigns searching for Pop~III stars at lower, more accessible, redshifts ($z\lesssim 5$), possibly even extending to recent times. If detected at lower redshifts, Pop~III systems could be studied in detail, directly measuring their IMF, which is out of reach at high redshifts, even with the next generation of telescopes, such as the {\it James Webb Space Telescope} (JWST). The challenge is to identify any such pockets of Pop~III star formation at more recent epochs, which would be extremely rare.

There are three main physical processes that regulate Pop~III star formation: (i) metal enrichment, (ii) Lyman-Werner (LW) radiation, and (iii) reionization. Locally, once the metallicity is above some critical value $Z_{\rm crit}\sim 10^{-6}-10^{-3.5}\ \rm Z_{\odot}$ (e.g. \citealt{bromm2003formation,omukai2005thermal,smith2009three}), star formation is shifted to the low-mass Population II/I (Pop~II/I) mode. This threshold is typically exceeded within atomic cooling haloes ($M_{\rm halo}\gtrsim 10^{7-8}\ \rm M_{\odot}$), which host the first galaxies (e.g. \citealt{wise2011birth,pawlik2013first,jeon2019signature}), and minihaloes externally enriched by nearby SNe \citep{wise2014birth,smith2015first,jeon2017}. However, metal enrichment is highly inhomogeneous, driven by complex interactions between SN blast waves and the ambient medium, accretion of primordial gas, and turbulent mixing during structure formation (e.g. \citealt{greif2010first,pan2013modeling,ritter2015metal}). As a result, even if the mean metallicity is above $Z_{\rm crit}$ at lower redshifts ($z\lesssim 10$), extremely metal-poor gas may still be available for Pop~III star formation, as implied by observed quasar spectra with little metal absorption (e.g. \citealt{simcoe2012extremely}).

Globally, LW radiation can photo-dissociate molecular coolants, thus delaying and relocating Pop~III star formation to occur in more massive haloes (e.g. \citealt{machacek2001simulations,oshea2006,safranek2012star,xu2013population}). Similarly, after reionization, star formation is significantly reduced in low-mass haloes ($M_{\rm halo}\lesssim 10^{9}\ \rm M_{\odot}$, e.g. \citealt{pawlik2015spatially,pawlik2017aurora,benitez2020detailed}), where hot ionized gas cannot collapse. In general, LW and ionizing photons can suppress and even terminate Pop~III star formation in low-mass haloes, but have little effect on massive haloes, where metal mixing is the key.

Taking into account all or some of these processes, Pop~III star formation at lower redshifts ($z\lesssim 10$) has been studied with semi-analytical models and cosmological simulations (e.g. \citealt{tornatore2007population,karlsson2008uncovering,trenti2009formation,muratov2013revisiting,pallottini2014simulating,jaacks2019legacy}). While semi-analytical models predict the termination of Pop~III star formation at $z\sim 5-16$ \citep{scannapieco2003detectability,yoshida2004era,greif2006two,hartwig2015,mebane2018persistence,chatterjee2020}, sharp cut-offs of Pop~III star formation are not seen in simulations for $z\gtrsim 2.5$, where the inhomogeneous nature of metal enrichment is better captured. High-resolution simulations, which at least marginally resolve minihaloes, generally agree that Pop~III star formation continues down to $z\sim 4-8$ at a level of a few $ 10^{-6}$ to $10^{-4}\ \rm M_{\odot}\ yr^{-1}\ Mpc^{-3}$, without a sharp decrease towards lower redshifts \citep{wise2011birth,johnson2013first,xu2016late,sarmento2018following}. Nevertheless, different sub-grid models for star formation, stellar feedback, and metal mixing are adopted with simplifying assumptions, leading to uncertainties and discrepancies in the detailed histories and environments of Pop~III star formation. Besides, it is interesting to investigate Pop~III star formation at even lower redshifts ($z\lesssim 4$), when it may finally end. However, it is still computationally prohibitive to run a cosmological hydrodynamics simulation that can well resolve minihaloes ($\lesssim 5\times 10^{4}\ \rm M_{\odot}$ for dark matter) down to $z=0$ in a representative volume ($V_{\rm C}\gtrsim 10^{6}\ \rm Mpc^{3}$ for $\nu\lesssim 2$ peaks). In this regime, we have to rely on extrapolation of what is learned in simulations at higher redshifts, together with semi-analytical techniques.

In light of this, we construct a theoretical framework of Pop~III star formation in the post-reionization epoch ($z\lesssim 6$), by combining simulation data and semi-analytical modelling of turbulent metal mixing and the reionization process, which may not be fully captured in simulations. Although we cannot provide a definitive answer to the question of when Pop~III star formation was terminated, our work serves as a flexible platform to address this challenge. Specifically, individual elements of the framework, such as the metal mixing efficiency and ionization history, can be constrained by separate studies. Among them are high-resolution, zoom-in simulations, focusing on metal transport in the wake of SN feedback. Furthermore, our approach can be applied to different cosmological simulations. It also enables us to explore the observational signatures of possible late Pop~III star formation at more recent cosmic times.

The paper is structured as follows. Section~\ref{s2} briefly describes the sub-grid models and setup of our cosmological simulations, for which details are available in \citealt{liu2020} (LB20, henceforth). In Section~\ref{s3}, we demonstrate how the key feedback processes that regulate Pop~III star formation arise during the evolution of the universe. Section~\ref{s4} presents our framework of Pop~III star formation after reionization, including its possible termination and the corresponding observational signatures. We summarize our findings and conclusions in Section~\ref{s5}.

\section{Simulating Early Star Formation}
\label{s2}
Our cosmological simulations are conducted with the \textsc{gizmo} code \citep{hopkins2015new}, using the Lagrangian meshless finite-mass (MFM) hydro solver, with a number of neighbours $N_{\mathrm{ngb}}=32$, and the Tree+PM gravity solver from \textsc{gadget-3} \citep{springel2005cosmological}. The properties of baryons are computed with a non-equilibrium primordial chemistry and cooling network for 12 primordial species, supplemented by metal-line cooling of \CII, \OI, \SiII\ and \FeII\  \citep{jaacks2018baseline}. Sub-grid models for Pop~III and Pop~II star formation, stellar feedback, black hole formation, accretion, dynamics, feedback and reionization are employed (LB20). Specifically, Pop~III and Pop~II are distinguished by a threshold metallicity $Z_{\rm th}=10^{-4}\ \rm Z_{\odot}$ ($\rm Z_{\odot}=0.02$). The former is characterized by a modified Larson IMF, $dN/dM\propto M^{-\alpha}\exp(-M^{2}_{\mathrm{cut}}/M^{2})$, with $\alpha=0.17$ and $M^{2}_{\mathrm{cut}}=20\ \mathrm{M}_{\odot}^{2}$ in the mass range $1-150\ \mathrm{M}_{\odot}$ \citep{jaacks2018baseline}, while the latter by a Chabrier IMF in the mass range $0.08-100\ \mathrm{M}_{\odot}$ \citep{jaacks2019legacy}. LW radiation is modelled with a background term derived from the simulated star formation rate density (SFRD) and a local term under the optically thin assumption, taking into account self-shielding for photo-dissociation. Ionization and SN feedback from massive stars are implemented in terms of their long-term (`legacy') thermal and chemical impact \citep{jaacks2018baseline,jaacks2019legacy}, as well as SN-driven winds (\citealt{springel2003wind}). Reionization is modelled with UV background (UVB) heating based on the photo-ionization rate from \citet{faucher2009new}, assuming a characteristic scale of $1\ \rm kpc$ for shielding in the intergalactic medium (IGM). Table~\ref{t1} summarizes the key parameters of SN feedback. We refer the reader to LB20 for detailed descriptions and calibrations of the sub-grid models.

\begin{table}
    \centering
    \caption{SN feedback parameters (for individual stellar particles with a fixed mass $m_{\star}\simeq 600\ \rm M_{\odot}$). $E_{\rm SN}$ is the total SN energy injected into the surrounding medium, $M_{Z}$ is the total mass of metals released, $t_{\star}$ is the typical lifetime of massive stars, i.e. the time interval between the star formation and SN events, and $r_{\rm final}$ is the final enrichment radius. Following \citet{jaacks2018baseline,jaacks2019legacy}, each Pop~III stellar particle embodies a few stars sampled from the input IMF on-the-fly, whose contributions are summed up to give the total SN energy and metal mass. Both core-collapse SNe (CCSNe; for stars in the mass range $M\sim 8-40\ \rm M_{\odot}$) and pair-instability SNe (PISNe; for $M\sim 140-150\ \rm M_{\odot}$) are considered, with $E_{\rm CCSN}=10^{51}\ \rm erg$, $y_{\rm CCSN}=0.1$, and $E_{\rm PISN}=10^{52}\ \rm erg$, $y_{\rm PISN}=0.5$, respectively \citep{karlsson2013pregalactic,nomoto2013nucleosynthesis}. In this way, the SN parameters actually vary from particle to particle. Here typical values (based on IMF average) are shown for illustration. While for Pop~II, we assume that all Pop~II stellar particles are identical, characterized by pre-calculated physical quantities (per unit stellar mass), i.e. $\langle E_{\rm SN}\rangle\equiv E_{\rm CCSN}\langle N_{\rm CCSN}\rangle$ 
    and $\langle y_{Z}\rangle \equiv y_{\rm CCSN}\langle f_{\rm CCSN}\rangle$, 
    given the average number $\langle N_{\rm CCSN}\rangle\simeq 0.011\ \rm M_{\odot}^{-1}$ and mass fraction $\langle f_{\rm CCSN}\rangle\simeq 0.17$ of CCSNe from integrations of the adopted Chabrier IMF. }
    \begin{tabular}{ccccc}
    \hline
        Type & $E_{\rm SN}$ & $M_{Z}$ & $t_{\star}$ & $r_{\rm final}$ \\
    \hline
        Pop~III &  $\sum_{i}N_{\star,i}E_{i}$ & $\sum_{i}y_{i}M_{\star,i}$ & 3~Myr & $\propto E_{\rm SN}^{0.383}$\\
         & $\sim 7\times 10^{51}\ \rm erg$ & $\sim 39\ \rm M_{\odot}$ & & $\sim 650\ \rm pc$\\
        \\
        Pop~II & $\langle E_{\rm SN}\rangle m_{\star}$ & $\langle y_{Z}\rangle m_{\star}$ & 10~Myr & $\propto E_{\rm SN}^{0.383}$\\
         & $\simeq 6.7\times 10^{51}\ \rm erg$ & $\simeq 10\ \rm M_{\odot}$ & & $\simeq 640\ \rm pc$\\
    \hline
    \end{tabular}
    \label{t1}
\end{table}

In this work, we focus on the most representative run \texttt{FDbox\_Lseed} in LB20 (referred as the/our simulation, henceforth), where the simulated region is a periodic cubic box of side length $L=4\ h^{-1}\rm Mpc$. The initial positions and velocities of simulation particles are generated with the \textsc{music} code \citep{hahn2011multi} at an initial redshift $z_{i}=99$ for the \textit{Planck} $\Lambda$CDM cosmology \citep{planck}: $\Omega_{m}=0.315$, $\Omega_{b}=0.048$, $\sigma_{8}=0.829$, $n_{s}=0.966$, and $h=0.6774$. The chemical abundances are initialized with the results in \citet{galli2013dawn}, following \citealt{liu2019global} (see their Table~1). The (initial) mass of gas (dark matter) particles is $9.4\times 10^{3}\ \rm M_{\odot}$ ($5.2\times 10^{4}\ \rm M_{\odot}$). The basic unit of star formation (i.e. the mass of stellar particles) is set to $m_{\star}\simeq 600\ \rm M_{\odot}$, for both Pop~III and Pop~II stars. This choice reflects the typical mass ($500-1000\ \mathrm{M}_{\odot}$) of a Pop~III star cluster, based on high-resolution simulations of Pop~III star formation in individual minihaloes (e.g. \citealt{stacy2013constraining,susa2014mass,machida2015accretion,stacy2016building,hirano2017formation,hosokawa2020}), and constraints from the observed global 21-cm absorption signal \citep{schauer2019constraining}. We have verified that the choice of $m_{\star}$ has little impact on processes involving Pop~II stars. The simulation stops at $z=4$, when the simulation volume is marginally representative for all $\nu\lesssim 2$ peaks. The simulation data are analysed with \textsc{yt}\footnote{\url{https://yt-project.org/doc/index.html}} \citep{turk2010yt}, and dark matter haloes are identified by the 
\textsc{rockstar}\footnote{\url{https://bitbucket.org/gfcstanford/rockstar/src}} halo finder \citep{behroozi2012rockstar}. 

\section{Key Feedback Effects}
\label{s3}

In this section, we discuss the build-up of global LW and ionizing radiation fields as well as metal enrichment in the simulation, which are the key feedback effects  for regulating Pop~III star formation. We also evaluate the uncertainties in our simulation results, by comparison with other simulations and observational constraints. In the next section, we introduce semi-analytical corrections to account for potentially underestimated feedback effects.

\subsection{Radiation feedback}
\label{s3.1}

\begin{figure*}
\hspace{-5pt}
\centering
\subfloat[LW]{\includegraphics[width= 1.065\columnwidth]{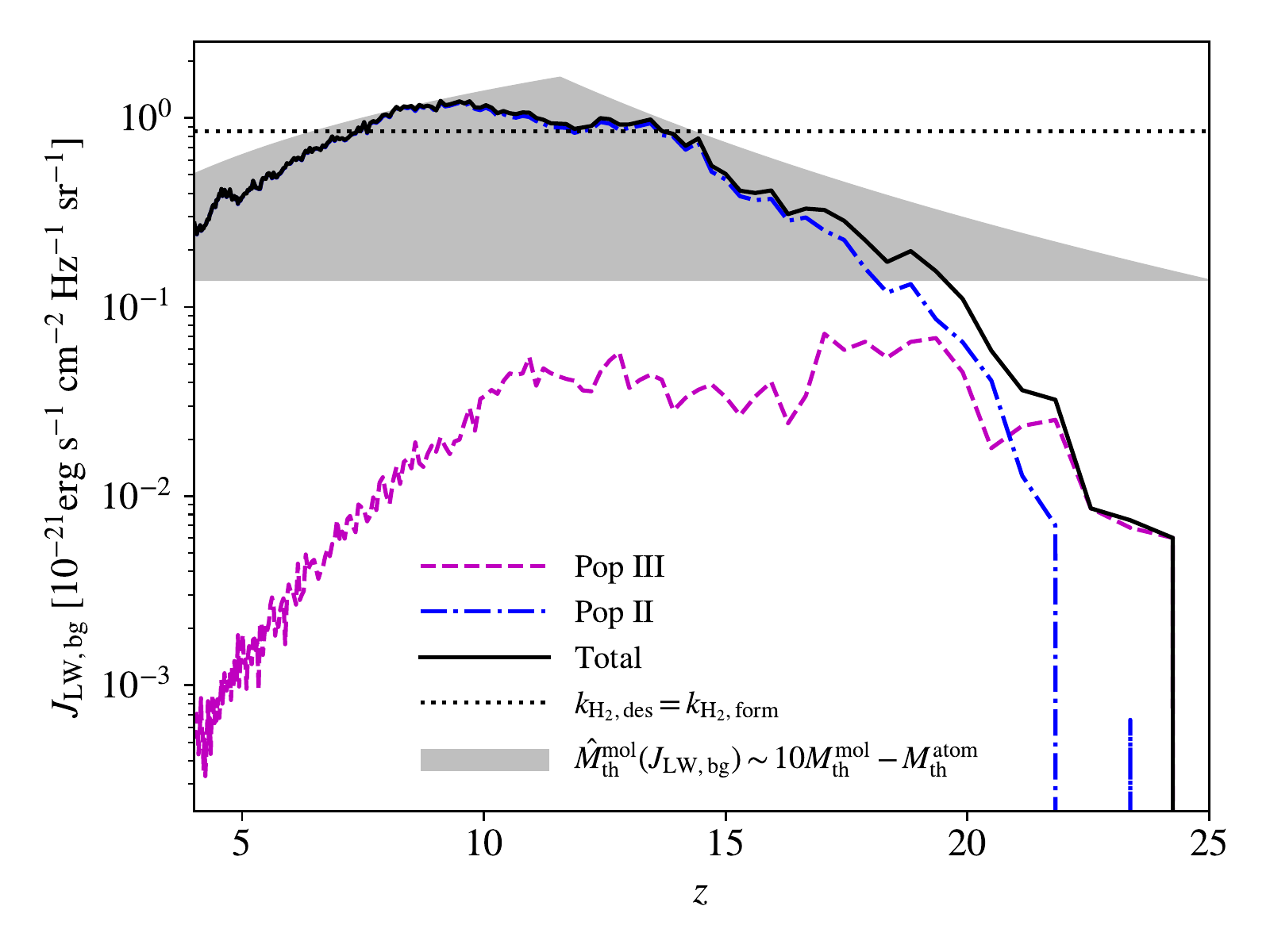}}
\subfloat[Ionization]{\includegraphics[width= 1.065\columnwidth]{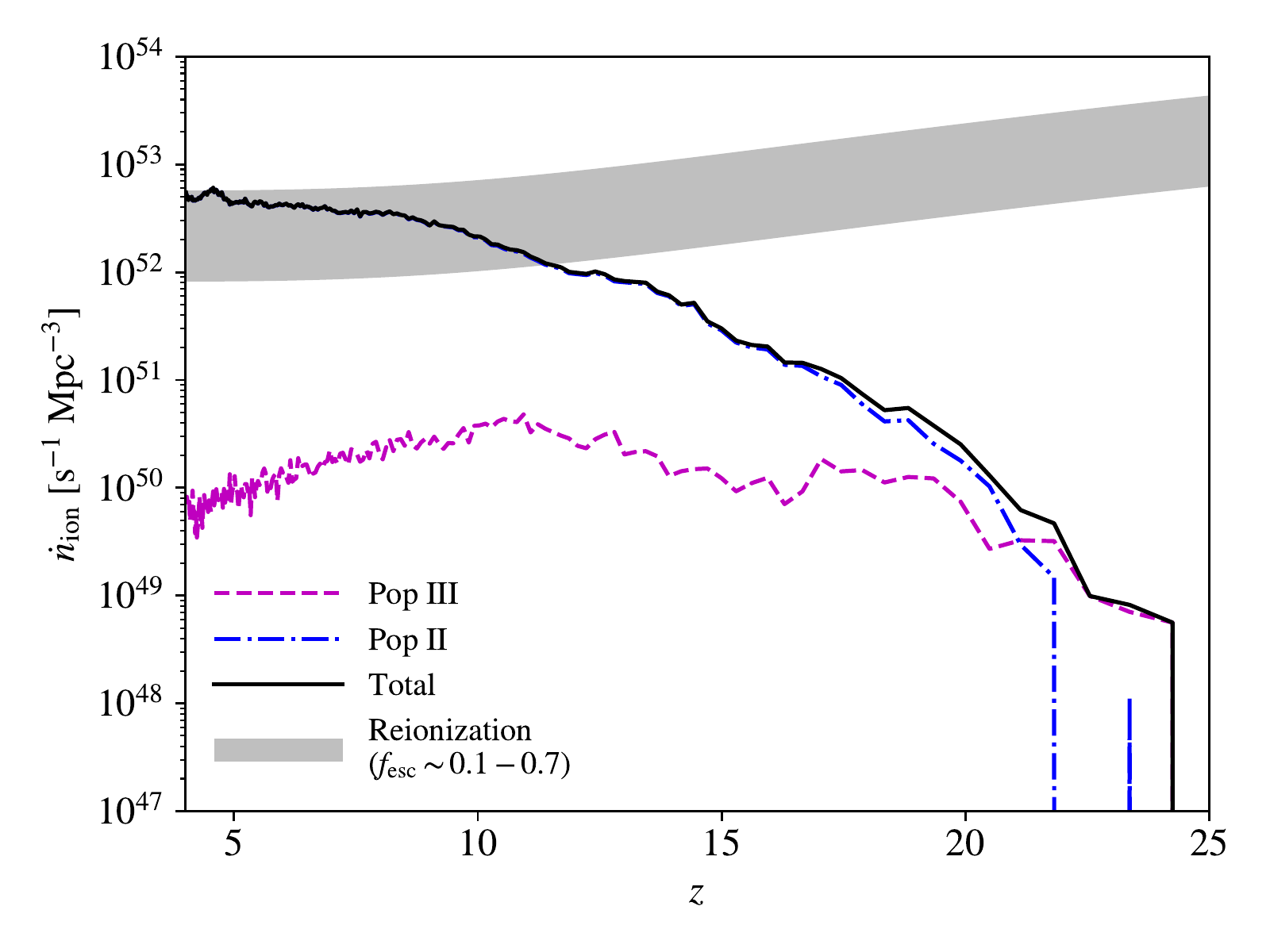}}
\vspace{-10pt}
\caption{Evolution of cosmic radiation backgrounds powered by all (solid), Pop~III (dashed) and Pop~II (dashed-dotted) stars. Left panel (a): LW background intensity (physical). For comparison, the `critical' intensity for significant suppression of Pop~III star formation is shown with the shaded region and the dotted horizontal line. The former is defined with the increase of the threshold halo mass for Pop~III star formation by a factor in the range of $10$ to $M_{\rm th}^{\rm atom}/M_{\rm th}^{\rm mol}$, inferred from the fit formula in \citet{machacek2001simulations,fialkov2014}. 
Here, $M_{\rm th}^{\rm mol}$ and $M_{\rm th}^{\rm atom}$ are the molecular and atomic cooling threshold masses \textit{without} LW feedback, respectively. 
The latter is defined with the $\rm H_{2}$ formation and destruction balance at a typical state of collapsing primordial gas (see main text). 
Right panel (b): production rate (co-moving) density of ionizing photons. The production rate required to fully ionize the IGM is shown with the shaded region (see equ.~20 of \citealt{johnson2013first}), given a typical range of escape fraction $f_{\rm esc}\sim 0.1-0.7$ \citep{so2014fully,paardekooper2015first}, and clumping factor evolution from \citealt{chen2020scorch} (for $\Delta<200$, see their equ.~13 and table.~2). }
\label{radbg}
\end{figure*}

Fig.~\ref{radbg} shows the evolution of cosmic radiation backgrounds, in terms of LW intensity $J_{\rm LW,bg}$ (left) and production rate density of ionizing photons $\dot{n}_{\rm ion}$ (right). Individual contributions from Pop~III and Pop~II stars are also shown. For LW radiation (see Sec.~2.2.2 of LB20), the Pop~II component exceeds that from Pop~III at $z\sim 20$ shortly after the onset of Pop~II star formation ($z\sim 23$), and dominates the LW background for $z\lesssim 15$. The fractional contribution from Pop~III decreases with decreasing redshift, and becomes negligible ($\lesssim 1$\%) in the post-reionization era ($z\lesssim 6$). 

To evaluate the strength of the LW feedback, we also plot the `critical' LW intensity for significant suppression of Pop~III star formation in low-mass haloes. Two definitions are employed for this `critical' intensity. The first is based on the increase of threshold halo mass for star formation caused by LW feedback, given by the fitting formula \citep{machacek2001simulations,fialkov2014}
\begin{align}
    \hat{M}_{\rm th}^{\rm mol}(J_{\rm LW})=M_{\rm th}^{\rm mol}\left[1+6.96(4\pi J_{\rm LW,21})^{0.47}\right]\mbox{\ ,}
\end{align}
where $\hat{M}_{\rm th}^{\rm mol}$ and $M_{\rm th}^{\rm mol}$ are the mass thresholds with and without LW feedback, $J_{\rm LW,21}=J_{\rm LW}/(10^{-21}\rm erg\ s^{-1}\ cm^{-2}\ Hz^{-1}\ sr^{-1})$, and we regard an increase by a factor in the range of $10$ to $M_{\rm th}^{\rm atom}/M_{\rm th}^{\rm mol}$ as `critical' (shaded region). Here we adopt the threshold masses $M_{\rm th}^{\rm mol}$ and $M_{\rm th}^{\rm atom}$ in \citet{trenti2009formation}, and further impose a lower limit of $10^{6}\ \rm M_{\odot}$ for $M_{\rm th}^{\rm mol}$, based on the simulations of \citet{Anna2018} for the typical case of $1\sigma$ baryon-dark-matter streaming velocity. 
The second is based on the $\rm H_{2}$ formation and destruction balance\footnote{We have taken into account self-shielding against LW radiation in calculating the $\rm H_{2}$ destruction rate, with the same method used in the simulation (see Sec.~2.2.2 of LB20), based on \citet{wolcott2011photodissociation}.} at a typical state of collapsing primordial gas with a temperature $T=250\ \rm K$, a density $n=100\ \rm cm^{-3}$, and an electron abundance $x_{\rm e}=10^{-5}$ (dotted horizontal line). It turns out that $J_{\rm LW,bg,21}\sim 0.1-1$ at $z\lesssim 19$, residing in the `critical' range given by the first definition, and exceeding the `critical' value according to the second definition at $z\sim 13-7$ (with $J_{\rm LW,bg,21}\sim 1$). In Sec.~\ref{s4.1}, we will show that Pop~III star formation is indeed shifted to more massive haloes (i.e. $M_{\rm halo}\gtrsim M_{\rm th}^{\rm atom}$) at $z\sim 13-7$. 

For ionizing radiation, we derive the total production rate density of ionizing photons as the summation of Pop~III and Pop~II contributions:
\begin{align}
    \dot{n}_{\rm ion}&=\dot{n}_{\rm ion,PopIII} + \dot{n}_{\rm ion,PopII}\ ,\\
    \dot{n}_{\mathrm{ion},k}&=\dot{\rho}_{\star, k}t_{\star,k}\langle \dot{N}_{\mathrm{ion},k}\rangle\ ,\quad k=\rm PopIII,\ PopII \ ,
\end{align}
where $\dot{\rho}_{\star}$ is the simulated SFRD, $t_{\star}$ the lifetime of massive stars (3 Myr for Pop~III and 10 Myr for Pop~II), and $\langle \dot{N}_{\mathrm{ion}}\rangle$ the luminosity of ionizing photons per unit stellar mass. Following \citet{visbal2015,schauer2019constraining}, we adopt $\langle \dot{N}_{\mathrm{ion}}\rangle\sim 10^{48}\ \rm s^{-1}\ M_{\odot}^{-1}$ for Pop~III, based on \citealt{schaerer2002properties} (see their table~4), and $\langle \dot{N}_{\mathrm{ion}}\rangle\sim 10^{47}\ \rm s^{-1}\ M_{\odot}^{-1}$ for Pop~II, based on \citealt{samui2007probing} (see their table~1), given a typical metallicity $Z\sim 0.05\ \rm Z_{\odot}$ (see Fig.~\ref{metal1}). Similar to the case of LW radiation, the Pop~II contribution starts to dominate at $z\sim 20$, and that of Pop~III drops below 1\% at $z\lesssim 10$. For comparison, we estimate the production rate density required for reionization, based on the `critical' SFRD \citep{johnson2013first} 
\begin{align}
    \dot{\rho}_{\star}&=0.05\ \mathrm{M_{\odot}\ yr^{-1}\ Mpc^{-3}}\notag\\ &\times\left(\frac{C}{6}\right)\left(\frac{f_{\rm esc}}{0.1}\right)^{-1}\left(\frac{1+z}{7}\right)^{3}\ ,
\end{align}
where $C$ is the clumping factor, and $f_{\rm esc}$ is the escape fraction. We consider the typical range of escape fraction $f_{\rm esc}\sim 0.1-0.7$ \citep{so2014fully,paardekooper2015first}, and clumping factor evolution from \citealt{chen2020scorch} (for $\Delta<200$, see their equ.~13 and table.~2). The `critical' production rate is reached in our simulation at $z\sim 10-4$, which is consistent with our implementation of the UVB heating and the resulting ionization history.

\begin{figure}
\includegraphics[width=1\columnwidth]{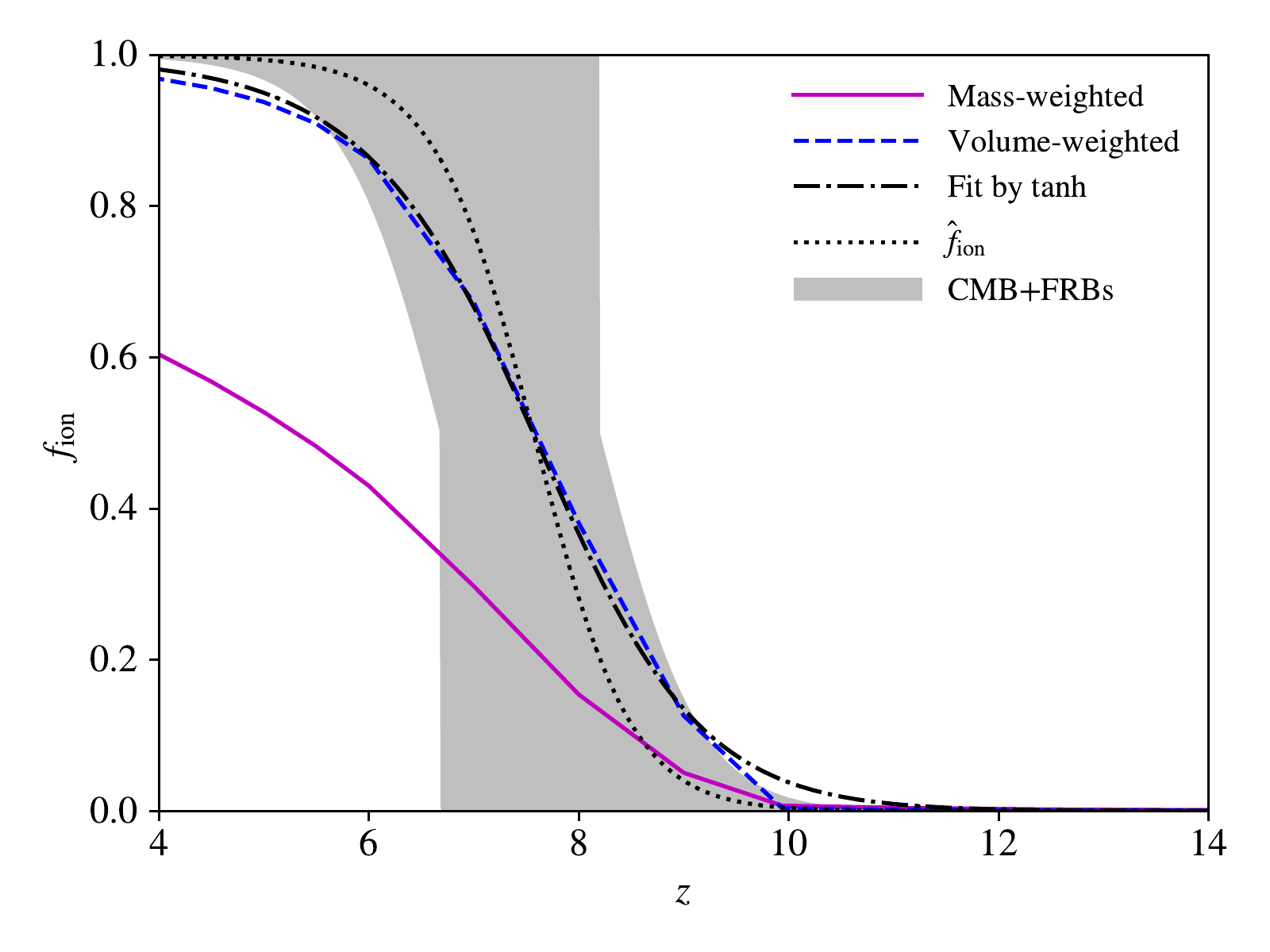}
\vspace{-20pt}
\caption{Evolution of mass- (solid) and volume- (dashed) weighted ionized fractions of hydrogen. The volume-weighted ionized fraction is fitted with a  tanh function (Equ.~\ref{e10}), 
which has two free parameters: the location $z_{\rm re}$ and width $\Delta z$ of reionization. The best-fit parameters are $z_{\rm re}\simeq 7.6$ and $\Delta z\simeq 1.6$ in our case (dashed-dotted). To correct for overestimated self-shielding, $\Delta z=0.94$ is adopted in further applications of the best-fit ionized fraction $\hat{f}_{\rm ion}$ (dotted), based on observational CMB and FRB constraints \citep{dai2020}, shown with the shaded region.}
\label{fion}
\end{figure}

This is demonstrated in Fig.~\ref{fion}, which shows the mass- and volume-weighted hydrogen ionized fractions in the simulation. Both fractions start to rise at $z\sim 10$, reaching 0.6 (mass-weighted) and 0.97 (volume-weighted) at the end of the simulation ($z=4$). We further fit the volume-weighted ionized fraction to the widely used tanh form \citep{lewis2008} 
\begin{align}
    \hat{f}_{\rm ion}=\frac{1}{2}(1-x_{\rm e}^{\rm rec})\left[1+\tanh\left(\frac{y_{\rm re}-y}{\Delta y}\right)\right]+x_{\rm e}^{\rm rec}\ ,\label{e10}
\end{align}
where $x_{\rm e}^{\rm rec}$ is the ionized fraction left over from recombination, $y(z)=(1+z)^{3/2}$, $\Delta y=1.5\sqrt{1+z_{\rm re}}\Delta z$, $z_{\rm re}$ is the redshift at which $\hat{f}_{\rm ion}=0.5$, and $\Delta z$ describes the duration of reionization. We keep $x_{\rm e}^{\rm rec}=2\times 10^{-4}$ fixed, and fit for $z_{\rm re}$ and $\Delta z$. As shown in Fig.~\ref{fion}, the fit is excellent at $z\lesssim 9$, with best-fit parameters $z_{\rm re}\simeq 7.6$ and $\Delta z\simeq 1.6$. 

In our simulation, reionization is delayed compared to what is inferred from \textit{Planck} data, where $z_{\rm re}=8.8_{-1.4}^{+1.7}$ \citep{planck}, but the difference is still within $1\sigma$. This implies that we may underestimate the effect of reionization feedback, as is also evident in the mass-weighted ionized fraction, which is only 0.6 at $z=4$. Note that the mass fraction of collapsed objects (i.e. haloes, with overdensities $\Delta\gtrsim 200$) is $\sim 0.1<1-f_{\rm ion}$ at $z\sim 4$, indicating that self-shielding in the IGM is significant. Indeed, we will show in Sec.~\ref{s4.1} that in our simulation the effect of reionization on star formation becomes important only at $z\lesssim 4.5$. This is unphysically late, likely caused by this overestimation of self-shielding against external UV radiation. Actually, the characteristic scale of $1\ \rm kpc$ assumed for IGM self-shielding leads to an optical depth $\tau_{\rm ion}>10$ for $\Delta\gtrsim 5$ (i.e. structures that have passed turnover in the spherical collapse model) at $z\gtrsim 3$. Therefore, we conclude that the effect of reionization is not fully captured in our simulation, and should be modelled separately (see Sec.~\ref{s4.2}). Meanwhile, to correct for the overestimated self-shielding, we reset $\Delta z=0.94$ for further applications of $\hat{f}_{\rm ion}$, based on the combined constraints ($z_{\rm re}=7.44 \pm 0.76$, $\Delta z<0.94$) from the cosmic microwave background (CMB) and fast radio bursts (FRBs; \citealt{dai2020}).

\subsection{Metal enrichment}
\label{s3.2}

\begin{figure*}
\hspace{-5pt}
\centering
\subfloat[Mass-weighted average]{\includegraphics[width= 1.065\columnwidth]{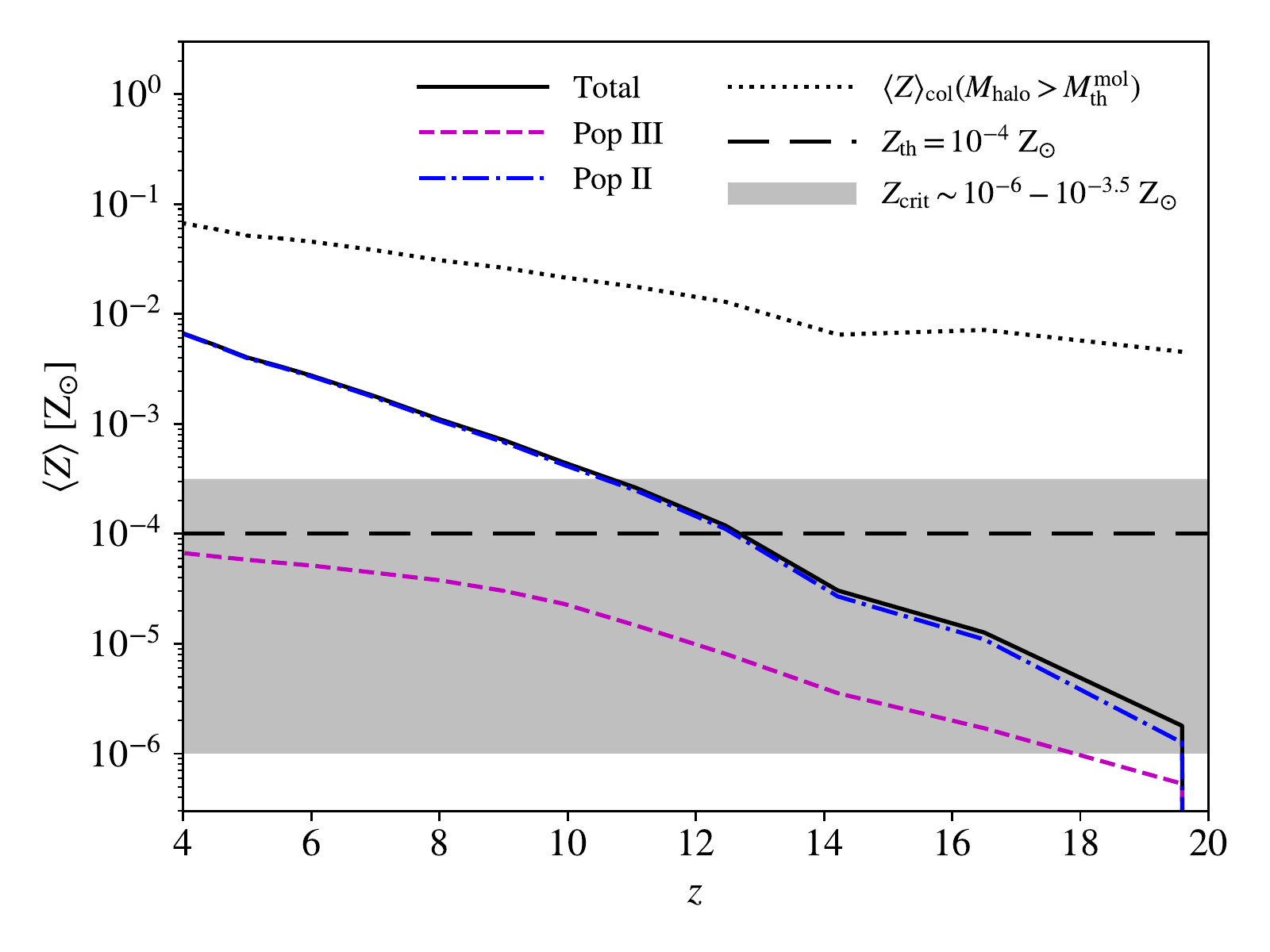}}
\subfloat[Volume filling fraction]{\includegraphics[width= 1.065\columnwidth]{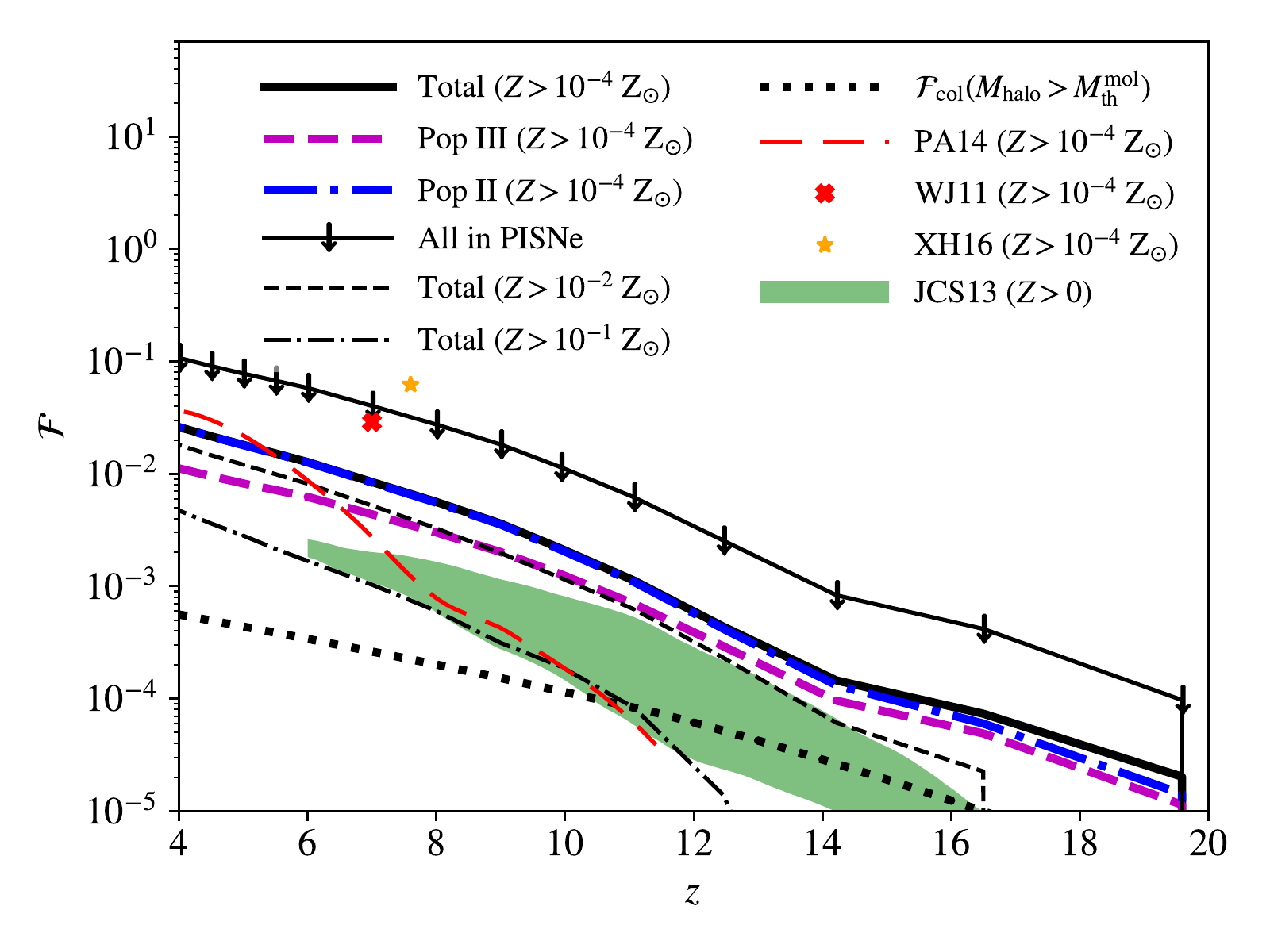}}
\vspace{-10pt}
\caption{Cosmic metal enrichment by all (solid), Pop~III (dashed) and Pop~II (dashed-dotted) SNe. Left panel (a): Mass-weighted mean metallicity of gas. The shaded region illustrates the `critical' metallicity for the Pop~III to Pop~II transition, within which the value adopted in our simulation is shown with the long dashed horizontal line. The dotted curve shows the mean metallicity in collapsed metal-enriched structures, i.e. haloes with $M_{\rm halo}>M_{\rm th}^{\rm mol}$, assuming that all metals are confined, i.e. $\langle Z\rangle_{\rm col}=\langle Z\rangle/f_{\rm col}$, where $f_{\rm col}$ is the mass fraction of such structures (see main text). 
Right panel (b): Volume filling fractions of gas at different metallicity levels. Results for $Z>10^{-4}\ \rm Z_{\odot}$ are shown with thick curves, while those for $Z>10^{-2}$ and $10^{-1}\ \rm Z_{\odot}$ with (normal) dashed and dashed-dotted curves, respectively. As an upper limit, we estimate the volume-filling fraction for the extreme case where all Pop~III stars end in PISNe, 
shown with the solid curve of downward arrows. For comparison, we plot the results from \citealt{wise2011birth} (WJ11), \citealt{johnson2013first} (JCS13), \citealt{pallottini2014simulating} (PA14) and \citealt{xu2016late} (XH16) with the cross, shaded region (which brackets the cases with and without LW feedback), long dashed curve and star, respectively. We also show the volume filling fraction of haloes with $M_{\rm halo}>M_{\rm th}^{\rm mol}$ (thick dotted), estimated as $\mathcal{F}_{\rm col}\simeq f_{\rm col}/200$. Here we use the data from the zoom-in run \texttt{FDzoom\_Hseed} (see LB20) for simplicity.}
\label{metal0}
\end{figure*}

Fig.~\ref{metal0} shows the global metal enrichment process in terms of the (mass-weighted) mean gas metallicity $\langle Z\rangle$ (left) and volume-filling fractions of gas $\mathcal{F}$ at different metallicity levels (right), derived from the zoom-in run \texttt{FDzoom\_Hseed} in LB20. The mean metallicity exceeds the critical metallicity for transitioning to Pop~II star formation, $Z_{\rm crit}\sim 10^{-6} - 10^{-3.5}\ \rm Z_{\odot}$ at $z\sim 20-10$. We further estimate the mean metallicity in collapsed metal-enriched structures, i.e. haloes with $M_{\rm halo}>M_{\rm th}^{\rm mol}$ (dotted curve), assuming that all metals are confined, i.e. $\langle Z\rangle_{\rm col}=\langle Z\rangle/f_{\rm col}$, where $f_{\rm col}=\int_{M_{\rm th}^{\rm mol}}^{\infty}n_{\rm h}(M)dM/\rho_{m}$ is the mass fraction of such structures, given the halo mass function $n_{\rm h}$ (calculated by \citealt{murray2013hmfcalc}), cosmic mean matter density $\rho_{m}$, and molecular cooling threshold $M_{\rm th}^{\rm mol}$ for star formation. The resulting mean metallicity $\langle Z\rangle_{\rm col}$ is always above $Z_{\rm crit}$ by at least one order of magnitude. Meanwhile, the volume-filling fraction remains below 10\%. This outcome reflects the inhomogeneous nature of metal enrichment, such that regions close to star formation sites are rapidly enriched, overshooting $Z_{\rm crit}$, while others remain extremely metal-poor \citep{scannapieco2003detectability}. Therefore, the volume filling fraction of significantly enriched gas $\mathcal{F}(Z>Z_{\rm th}=10^{-4}\ \rm Z_{\odot})$ is a better indicator of the effect of metal-enrichment on Pop~III star formation than the mean metallicity $\langle Z\rangle$. We also compare the volume filling fraction of metal-enriched mass to that of haloes with $M_{\rm halo}>M_{\rm th}^{\rm mol}$. The latter is estimated as $\mathcal{F}_{\rm col}\simeq f_{\rm col}/200$. We find that $\mathcal{F}_{\rm col}<\mathcal{F}$, even for $Z>10^{-1}\ \rm Z_{\odot}$ at $z\lesssim 11$, which is a sign of metal enrichment of the IGM driven by galactic outflows. Similar trends are also seen in previous studies \citep{wise2011birth,johnson2013first,pallottini2014simulating,xu2016late}.

Similar to radiation feedback, the Pop~III contribution to the mean metallicity of gas also decreases towards lower redshifts, dropping to $\sim 1$\% at $z=4$. However, the Pop~III contribution to the volume-filling fraction of significantly metal-enriched gas is always non-negligible (40\%-70\%). The reason is that Pop~III star formation tends to occur in low-density regions (away from previous star formation and metal enrichment activity) where SN bubbles can expand to larger volumes. Actually, in the extreme case where all Pop~III stars end in pair-instability SNe (PISNe), the volume-filling fraction (estimated by rescaling the Pop~III contribution based on the boost of SN energy by PISNe) will be dominated ($\gtrsim 90$\%) by Pop~III stars and reaches 10\% at $z=4$. This feature is also seen in the recent cosmological simulation from \citet{takanobu2020}, which finds that metals of Pop~III origin dominate in low-density regions ($\delta \lesssim 10$) at $z\sim 3$. Particularly, their models (b) and (c), which resemble our Pop~III IMF, predict a range of mean (Pop~III) metallicity $\sim 10^{-5}-10^{-3.5}\ \rm Z_{\odot}$ across regions of different overdensities (see their fig.~5), consistent with our result $\langle Z\rangle\simeq 7\times 10^{-5}\ \rm Z_{\odot}$ (corresponding to a mean metal mass density of $\sim 10^{-34}\ \rm g\ cm^{-3}$) at $z= 4$. 

We further compared our volume-filling fractions with literature results \citep{wise2011birth,johnson2013first,pallottini2014simulating,xu2016late}. For instance, our $\mathcal{F}(Z>10^{-4}\ \rm Z_{\odot})$ agrees well with that in \citet{pallottini2014simulating} at $z\lesssim 5$, but is higher by up to a factor of 10 at $z\gtrsim 6$ than those in \citet{johnson2013first,pallottini2014simulating}. Meanwhile, the simulations in \citet{wise2011birth,xu2016late} with the \textsc{enzo} code predict $\mathcal{F}(Z>10^{-4}\ \rm Z_{\odot})\sim 3-6$\% at $z\sim 7-8$, higher than our results by up to a factor of 10. 
Our extreme model, where all Pop~III stars end in PISNe, places an upper limit on the volume-filling fraction of 10\% at $z=4$, which is also consistent with the value 15\% from the corresponding model (a) in \citet{takanobu2020}. 
In general, the discrepancies in different simulations are significant (up to two orders of magnitude), reflecting the uncertainties in sub-grid models for SN feedback and metal transport. It remains uncertain to what extent the fine-grain metal mixing process (at a scale of $\sim 10^{-3}$~pc; \citealt{spitzer2006physics,sarmento2016following}) is captured in such cosmological simulations with limited resolution ($\Delta x\gtrsim 10$~pc). Therefore, the late-time ($z\lesssim 6$) Pop~III star formation in our simulation may not be adequately captured, due to the imperfect treatment of metal transport.

To be more specific, we assume that metals are instantaneously mixed into the gas particles enclosed by the final radius of SN shell expansion, while in reality, the timescale for complete mixing can be non-negligible ($\sim 1-10$~Myr) at the scale of gas particles ($m_{\rm gas}\sim 10^{4}\ \rm M_{\odot}$), marking the resolution limit of the simulation\footnote{According to \citet{pan2013modeling,sarmento2016following}, the dynamical mixing timescale is $\sim 1-10$~Myr, for our gas particles at a scale of $\Delta x\sim (m_{\rm gas}/\rho_{\rm gas})^{1/3}\sim 70$~pc, under the typical conditions in SN remnants with a mean matallicity $\bar{Z}\sim 10^{2}-10^{4}\ Z_{\rm crit}$, a density $n\sim 10^{-3}-1\ \rm cm^{-3}$, a temperature $T\sim 2\times 10^{4}$~K and a Mach number $\mathcal{M}\sim 1$.}. That is to say, even if the mean metallicity of a gas particle is above $Z_{\rm crit}$, there could still be a significant fraction of gas with $Z<Z_{\rm crit}$. It is shown in \citet{sarmento2016following,sarmento2018following} that taking into account such unresolved metal-poor gas will enhance the Pop~III SFRD by a factor of 2. On the other hand, we do not smooth gas metallicities with any kernels, such that the metallicity of a gas particle is only affected by nearby SNe. Our approach is valid at the scale of $m_{\rm gas}\sim 10^{4}\ \rm M_{\odot}$, as the MFM method sets the mass fluxes across mesh boundaries explicitly to zero. However, turbulent metal diffusion at unresolved scales across different gas particles is not captured, which may reduce the fraction of metal-poor gas in a halo. Besides, the mesh itself in \textsc{gizmo} is actually smoothed. Therefore, in Sec.~\ref{s4.2}, we use semi-analytical models of metal-mixing to correct for the potentially overestimated amount of metal-poor gas in the simulated Pop~III SFRD. It is also possible to implement passive scalar diffusion in simulations (see \citealt{hopkins2017anisotropic}). We defer such investigations to future work.

\begin{figure}
\includegraphics[width=1.\columnwidth]{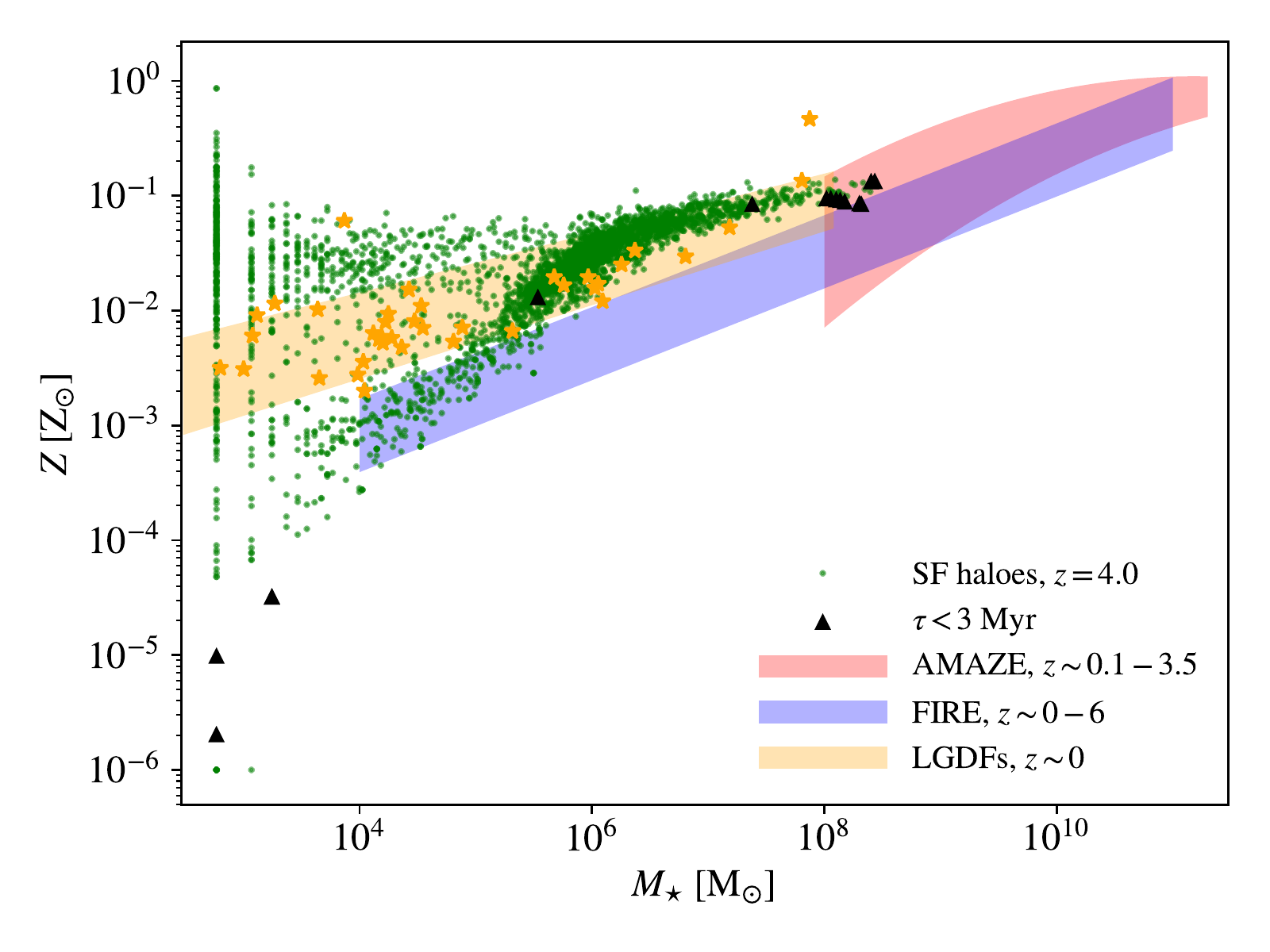}
\vspace{-20pt}
\caption{Stellar metallicity-mass relation. Simulated star-forming haloes at $z=4$ are shown with green dots, and the ones undergoing recent Pop~III star formation within 3 Myr are labelled with triangles. Systems with $Z< 10^{-6}\ \rm Z_{\odot}$ are plotted at $Z= 10^{-6}\ \rm Z_{\odot}$. 
For comparison, we also plot the results for FIRE simulation \citep{ma2016origin}, AMAZE observations \citep{maiolino2008amaze}, and local group dwarfs (LGDFs; \citealt{simon2019faintest}) with blue, red and orange shaded regions. Individual LGDFs are labelled with stars.}
\label{metal1}
\end{figure}

Although our treatment of metal transport is idealized, the observed stellar metallicity-mass relation \citep{maiolino2008amaze,simon2019faintest} is well reproduced by our simulation in a broad mass range $M_{\star}\sim 10^{3}-10^{8}\ \rm M_{\odot}$, especially for local group dwarfs (LGDFs), as shown in Fig.~\ref{metal1}. We find that the scatter in stellar metallicity increases with decreasing stellar mass, in particular for $M_{\star}\lesssim 5\times 10^{5}\ \rm M_{\odot}$, also consistent with the trend in observations. However, at a fixed stellar mass, the stellar metallicity is higher in our simulation compared with the FIRE simulations \citep{ma2016origin}, again demonstrating the uncertainties in cosmological simulations regarding metal enrichment.

\begin{figure}
\includegraphics[width=1\columnwidth]{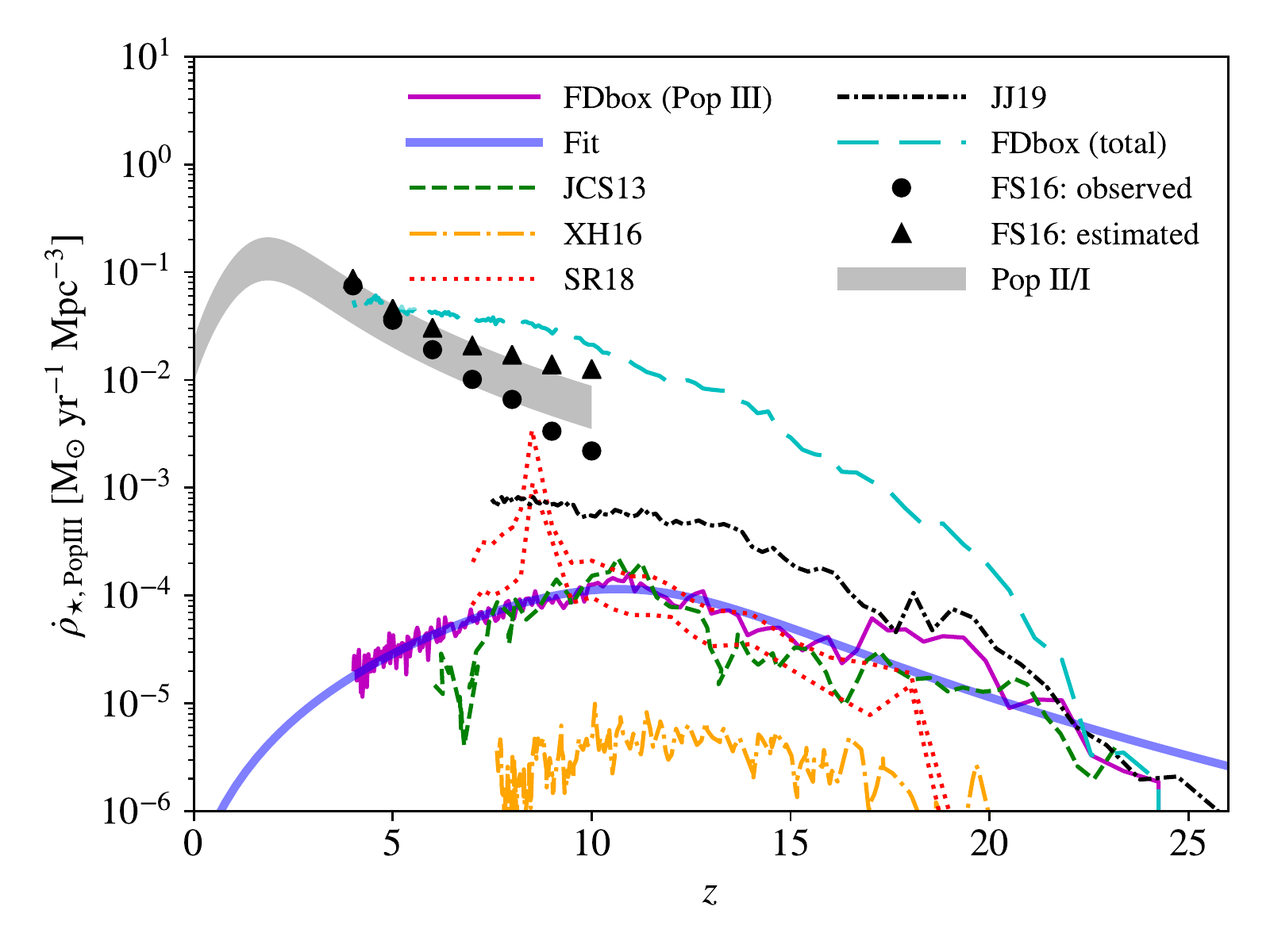}
\vspace{-20pt}
\caption{Co-moving Pop~III SFRD. The results from our fiducial run and the corresponding fit are shown with thin and thick solid curves. We further plot the results from other cosmological simulations in \citealt{johnson2013first} (JCS13; with LW feedback), \citealt{xu2016late} (XH16), \citealt{sarmento2018following} (SR18) and \citealt{jaacks2019legacy} (JJ19), with the dashed, dashed-dotted, dotted and densely dashed-dotted curves, which demonstrate the range of Pop~III star formation histories in current models. Note that the XH16 results are based on a zoom-in simulation for a low-density region ($\langle\delta\rangle=-0.26$ at $z=8$), which should be regarded as lower limits. The SR18 results include two cases with (upper) and without (lower) unresolved inefficient metal mixing. For comparison, we plot the (extrapolated) Pop~II/I ($\approx$ total) SFRD $\dot{\rho}_{\star}=0.015(1+z)^{2.7}/\{1+[(1+z)/2.9]^{5.6}\}\ \rm M_{\odot}\ yr^{-1}\ Mpc^{-3}$ (with 0.2 dex scatters) from \citealt{madau2014araa} (shaded region), inferred by UV and IR galaxy surveys, such as \citealt{finkelstein2016observational} (FS16; data points). The corresponding simulated total SFRD is shown with the long-dashed curve.}
\label{sfrd}
\end{figure}

\section{Pop~III star formation after reionization}
\label{s4}

In this section, we demonstrate our framework for Pop~III star formation after reionization ($z\lesssim 6$). We first characterize a representative sample of simulated haloes with recent Pop~III star formation at $z\sim 4-6$, considering their mass and metallicity distributions, as well as the masses and locations of active Pop~III stars within them (Sec.~\ref{s4.1}). Based on this sample, we then employ semi-analytical models for metal mixing and reionization to extrapolate the Pop~III SFRD to $z=0$ (Sec.~\ref{s4.2}). Finally, we discuss the observational constraints and possible signatures of Pop~III star formation in the post-reionization epoch ($z\lesssim 6$), as well as its potential termination (Sec.~\ref{s4.3}).

The starting point of our framework is the simulated (co-moving) Pop~III SFRD, which is shown in Fig.~\ref{sfrd}, in comparison with literature results \citep{johnson2013first,xu2016late,sarmento2018following,jaacks2019legacy}. Our Pop~III SFRD peaks at $z\sim 10$ with $\sim 10^{-4}\ \rm M_{\odot}\ yr^{-1}\ Mpc^{-3}$, and drops to $\sim 2\times 10^{-5}\ \rm M_{\odot}\ yr^{-1}\ Mpc^{-3}$ at $z=4$. We fit the simulated Pop~III SFRD to the form \citep{madau2014araa}
\begin{align}
    \frac{\dot{\rho}_{\star,\mathrm{PopIII}}^{\rm sim}(z)}{\rm M_{\odot}\ yr^{-1}\ Mpc^{-3}}=\frac{a(1+z)^{b}}{1+[(1+z)/c]^{d}}\ ,\label{fit}
\end{align}
which leads to best-fit parameters $a=765.7$, $b=-5.92$, $c=12.83$ and $d=-8.55$. For the post-reionization epoch ($z\lesssim 6$), this is approximately equivalent to a power-law extrapolation $\propto (1+z)^{b-d}\simeq (1+z)^{2.6}$, as $d<0$ in our case. 
Interestingly, the power-law index here is similar to that of the Pop~II/I SFRD from \citet{madau2014araa}. Integrating $\dot{\rho}_{\star,\mathrm{PopIII}}^{\rm sim}(z)$ across cosmic history gives the density of all Pop~III stars ever formed, $\sim 10^{5}\ \rm M_{\odot}\ Mpc^{-3}$ (in which 55\% comes from $z>6$), consistent with the constraints in \citet{visbal2015}, set by \textit{Planck} data.

Our Pop~III SFRD agrees well with \citet{johnson2013first} at $z\gtrsim 7$ and \citet{sarmento2018following} at $z\gtrsim 9$, but is lower (higher) compared with that in \citealt{jaacks2019legacy} (\citealt{xu2016late}) at $z\gtrsim 7$. This can be explained with the fact that \citet{jaacks2019legacy} did not include mechanical SN feedback, while \citet{xu2016late} targeted a low-density region ($\langle\delta\rangle=-0.26$ at $z=8$), whose results should be regarded as lower limits. In general, our Pop~III SFRD is approximately the median value among various simulation results \citep{tornatore2007population,wise2011birth,johnson2013first,xu2016late,sarmento2018following}. In Fig.~\ref{sfrd}, we also plot the simulated total SFRD (dominated by Pop~II/I at $z\lesssim 22$), which is consistent with observations within a factor of 2 \citep{madau2014araa,finkelstein2016observational}. We refer the reader to Section 3 of LB20 for more detailed comparisons between our simulations and observations.

\subsection{Host haloes of Pop~III stars}
\label{s4.1}

\begin{figure}
\hspace{-5pt}
\centering
\includegraphics[width=1\columnwidth]{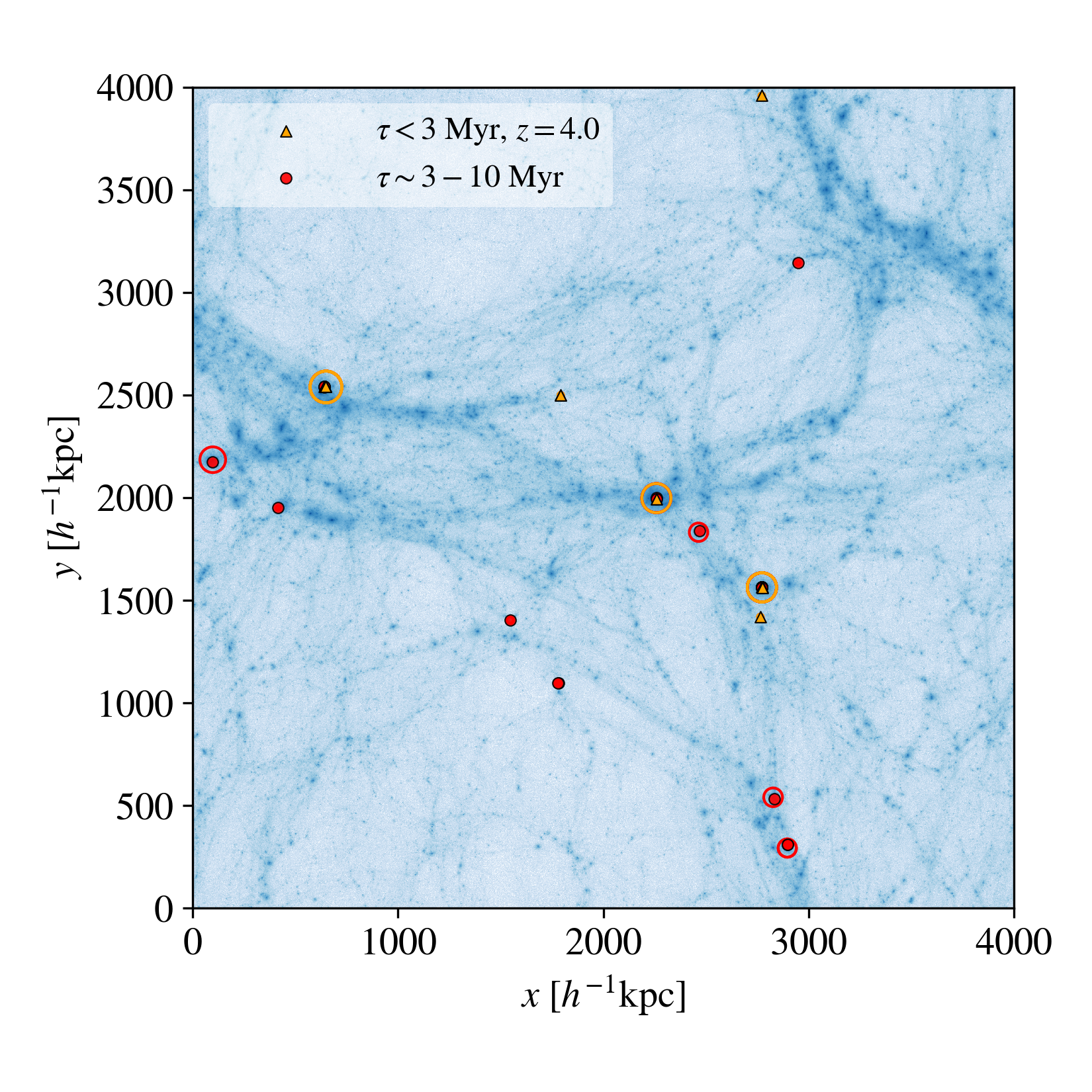}
\vspace{-20pt}
\caption{Cosmic web from the fiducial run at $z=4$, in terms of the projected distribution of dark matter (in co-moving coordinates, with a thickness of $4\ h^{-1}\rm Mpc$). Pop~III stellar particles with ages $\tau<3$ and $\sim 3-10$~Myr are labelled with orange triangles and red filled circles. Their host haloes are also shown with empty circles whose sizes reflect their virial radii. Note that small haloes ($M_{\rm halo}\lesssim 10^{10}\ \rm M_{\odot}$) have been covered by the labels of Pop~III stars.}
\label{popIIIdis}
\end{figure}

\begin{figure*}
\centering
\includegraphics[width=1.4\columnwidth]{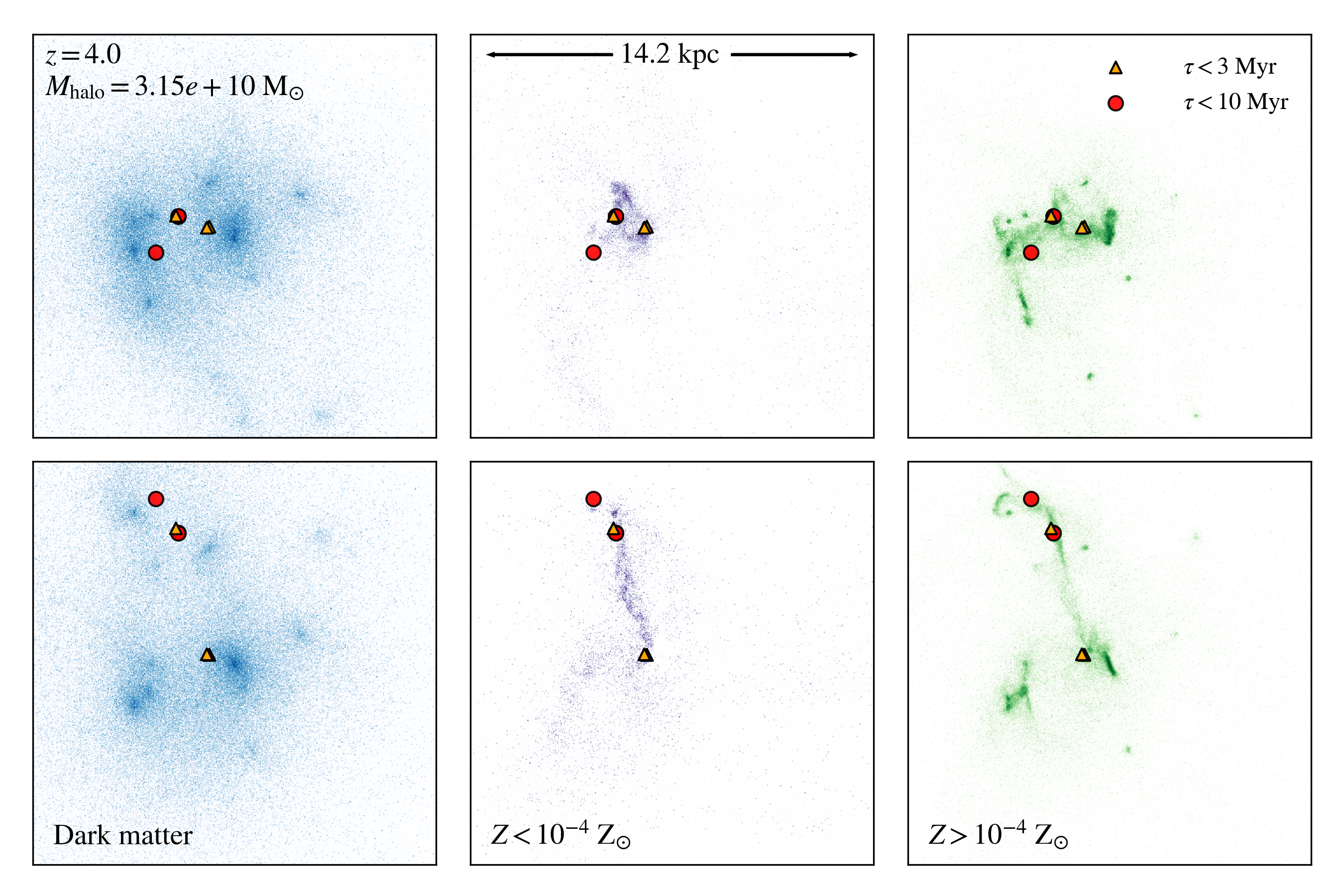}
\vspace{-10pt}
\caption{Edge-on ($xy$, top) and Face-on ($xz$, bottom) projected distributions of dark matter (left), metal-poor gas ($Z<10^{-4}\ \rm Z_{\odot}$, middle) and metal-enriched gas ($Z>10^{-4}\ \rm Z_{\odot}$, right) in one halo of $M_{\rm halo}\simeq 3\times 10^{10}\ \rm M_{\odot}$ with recent Pop~III star formation at $z=4$. Pop~III star particles with ages $\tau<3$ and $\sim 3-10$~Myr are labelled with orange triangles and red filled circles. The physical scale of the cubic region shown is $\sim 14\ \rm kpc$. It is evident that Pop~III stars tend to form at the edges of sub-structures, reminiscent of the `Pop~III wave' theory \citep{tornatore2007population}.}
\label{gasdis}
\end{figure*}

A fundamental question for Pop~III star formation at late times is where Pop~III stars could possibly continue to form. As an example, Fig.~\ref{popIIIdis} shows the locations of (active) Pop~III stellar particles with ages $\tau<3$~Myr and $\tau\sim 3-10$~Myr, on top of the cosmic web in the last simulation snapshot at $z=4$. We identify the host of a Pop~III particle as the most massive halo that encloses the Pop~III particle within its virial radius. Massive ($M_{\rm halo}\gtrsim 10^{10}\ \rm M_{\odot}$) host haloes of active Pop~III star formation are also shown in Fig.~\ref{popIIIdis}, constituting 50\% of the host halo population at $z=4$, thus indicating that formation of Pop~III stars in massive haloes is important. The reason is that metal mixing is inefficient in our simulation, such that metal-poor gas in dense filaments (i.e. cold accretion flows) can still form Pop~III stars, even though the densest regions within the halo have been significantly enriched by previous SNe. This process is illustrated in Fig.~\ref{gasdis}, where (active) Pop~III particles are plotted on top of the projected distributions of dark matter (left), metal-poor gas ($Z<10^{-4}\ \rm Z_{\odot}$, middle) and metal-enriched gas ($Z>10^{-4}\ \rm Z_{\odot}$, right), for one of the most massive haloes at $z=4$ that host active Pop~III stars with $M_{\rm halo}\simeq 3\times 10^{10}\ \rm M_{\odot}$. This halo is still under assembly with a few (groups) of sub-haloes separated by a few (physical) kpc, where Pop~III stars are formed on the edges of such sub-structures. Actually, Pop~III stars tend to form at the `connection points' of (metal-enriched) sub-structures and dense filaments rich in metal-poor gas. This trend is consistent with the `Pop~III wave' scenario \citep{tornatore2007population}, which is also seen in previous simulations (e.g. \citealt{pallottini2014simulating,xu2016late}). Besides, recent work by \citet{bennett2020} found that inflows of cold dense gas are significantly enhanced with better resolution of shocks, leading to metal-poor star formation in primordial filaments, for even more massive haloes ($M_{\rm halo}\sim 10^{12}\ \rm M_{\odot}$).

To characterize the host haloes of Pop~III stars after reionization ($z\lesssim 6$), we combine 4 snapshots at $z=4$, 4.5, 5 and $6$ to construct a sample of 145 (52) haloes that have recent Pop~III star formation within 10 (3) Myr (the representative sample, henceforth). As mentioned in Sec.~\ref{s3.1}, and to be further discussed below, reionization feedback is not well captured in our simulation, while the effect of LW feedback is treated more realistically. Therefore, the representative sample from our simulation effectively corresponds to the case under a moderate LW background, but \textit{without} reionization feedback. In the next subsection, additional corrections are made to fully take into account the effect of reionization.

We divide the Pop~III host haloes into three groups, based on the atomic cooling threshold $M_{\rm th}^{\rm atom}$ and the (dark matter+baryonic) Jeans mass of fully ionized gas
\begin{align}
    M_{\rm J,ion}&\simeq 6.7\times 10^{8}\ \mathrm{M_{\odot}}\notag\\
    &\times \left[\frac{(1+z)^{3}\Delta}{5^3\times 125}\right]^{-1/2}\left(\frac{T_{\rm b}}{20000\ \rm K}\right)^{3/2} \mbox{\ ,}\label{mj}
\end{align}
where $\Delta$ is the overdensity and $T_{\rm b}$ the temperature of ionized gas. We use the Jeans mass (Equ.~\ref{mj}) to approximate the halo mass threshold below which star formation is significantly suppressed due to reionization (i.e. the filtering mass, \citealt{gnedin2000effect}). We adopt $\Delta=125$ and $T_{\rm b}=20,000$~K in accordance with more complex calculations and simulations \citep{pawlik2015spatially,pawlik2017aurora,benitez2020detailed,hutter2020astraeus}. For simplicity, we evaluate $M_{\rm th}^{\rm atom}$, $M_{\rm J,ion}$ at $z=4$, and apply $M_{\rm th}^{\rm atom}\simeq 1.2\times 10^{8}\ \rm M_{\odot}$, $M_{\rm J,ion}\simeq 6.7\times 10^{8}\ \mathrm{M_{\odot}}$ to the entire representative sample at $z\sim 4-6$. 

\begin{figure}
\includegraphics[width=1\columnwidth]{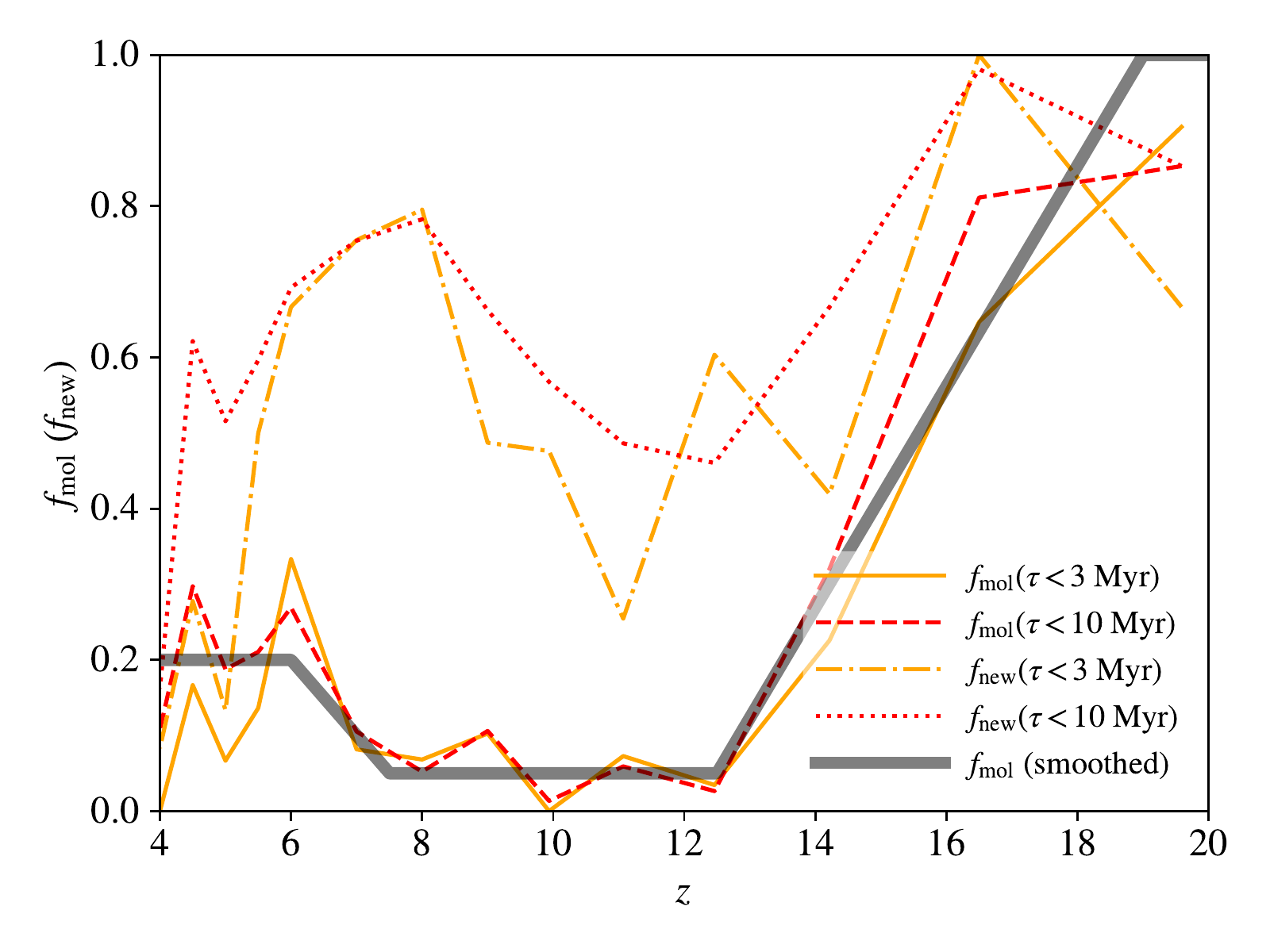}
\vspace{-20pt}
\caption{Fractions of active Pop~III stars in molecular cooling haloes ($f_{\rm mol}$) and new star-forming haloes ($f_{\rm new}$), for $\tau<3$ (solid and dashed-dotted) and 10 (dashed and dotted) Myr. 
We also show a smoothed version of $f_{\rm mol}$ with the thick gray curve, in which $f_{\rm mol}=0.2$ at $z<6$ is assumed to denote the case under a moderate LW background but \textit{without} reionization. The effect of reionization will be modelled separately in Sec.~\ref{s4.2} given the smoothed $f_{\rm mol}$ as a starting point.}
\label{fiso}
\end{figure}

The first group refers to the `classical' formation sites of Pop~III stars with $M_{\rm halo}<M_{\rm th}^{\rm atom}$, the minihaloes, where molecular (hydrogen) cooling dominates, and which are particularly important at high-$z$. This group itself is interesting, as it reflects how feedback regulates Pop~III star formation. In Fig.~\ref{fiso}, we plot the fraction of active Pop~III stars in molecular cooling haloes, $f_{\rm mol}$, in comparison with the fraction of newly star-forming haloes\footnote{A halo with active Pop~III stars is called a newly star-forming halo, if it has not experienced any star formation activities prior to the recent Pop~III star formation. }, $f_{\rm new}$, for $z\sim 4-20$ (i.e. isolated Pop~III star formation). In general, $f_{\rm new}> f_{\rm mol}$, especially for $z\lesssim 13$, which indicates that at lower redshifts, the majority of isolated Pop~III star formation occurs in atomic cooling haloes. $f_{\rm mol}$ drops from close to 1 to a few percent when $z$ decreases from $\sim 20$ to $\sim 13$, resulting from the suppression/delay of star formation in molecular cooling haloes by LW radiation. Actually, $f_{\rm mol}$ anti-correlates with the background LW intensity $J_{\rm LW,bg}$, shown in the left panel of Fig.~\ref{radbg}. For instance, $f_{\rm mol}$ remains a few percent at $z\sim 13-7$ when $J_{\rm LW,bg}$ is above the `critical' value ($J_{\rm LW,bg,21}\gtrsim 1$). Similar trends are also seen in the recent simulation of \citealt{danielle2020} (see their fig. 5). Both $f_{\rm new}$ and $f_{\rm mol}$ decrease rapidly at $z\lesssim 4.5$, where reionization starts to take effect. This is later than expected, for the reason explained in Sec.~\ref{s3.1}. In the next subsection, for the purpose of post-processing, we use a smoothed version of $f_{\rm mol}$, assuming that $f_{\rm mol}=0.2$ at $z<6$, which again reflects the case under a moderate LW background ($J_{\rm LW,bg,21}\sim 0.1-1$), but \textit{without} reionization feedback. 

\begin{figure}
\includegraphics[width=1\columnwidth]{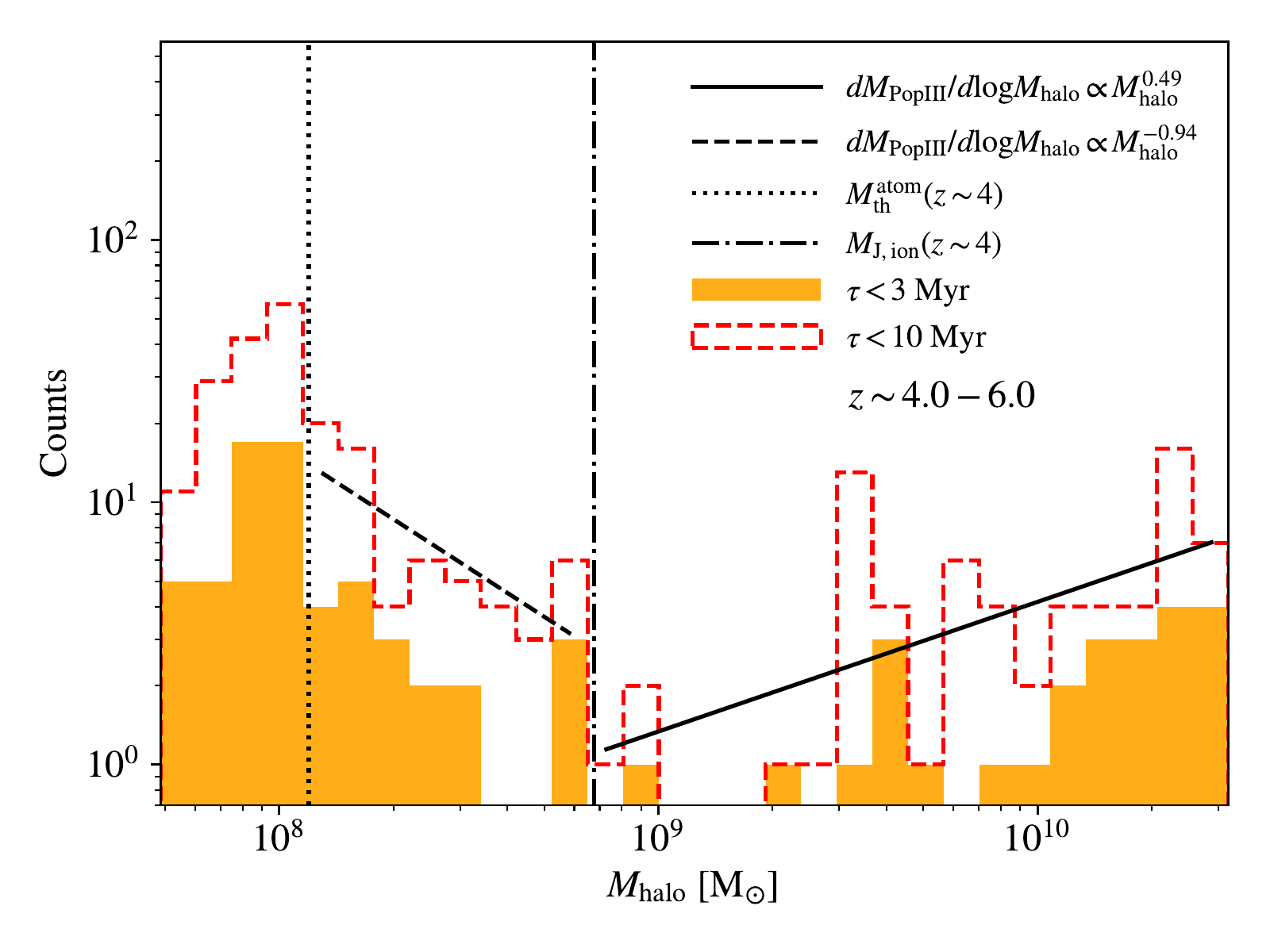}
\vspace{-20pt}
\caption{Halo mass distribution (in log scale) of the representative halo sample with recent Pop~III star formation within 3 (orange histograms) and 10 (red dashed contour) Myr. Haloes are weighted by enclosed masses of active Pop~III stars, such that the distribution here is proportional to $dM_{\rm PopIII}/d\log M_{\rm halo}$. The atomic cooling threshold $M_{\rm th}^{\rm atom}$ and Jeans mass of haloes with fully ionized gas $M_{\rm J,ion}$ are shown with the dotted and dashed-dotted vertical lines. The distribution at $M_{\rm halo}\gtrsim M_{\rm J,ion}$ can be approximated with a power-law of index $\alpha_{m}\sim 0.5$ (solid), while that at $M_{\rm th}^{\rm atom}\lesssim M_{\rm halo}\lesssim M_{\rm J,ion}$ can be described by another power-law with $\alpha_{m}\sim -1$. The total masses of active Pop~III stars in these two groups of haloes are almost identical (with $<10\%$ difference).}
\label{mhdis}
\end{figure}

The other two groups refer to haloes with $M_{\rm halo}\in [ M_{\rm th}^{\rm atom}, M_{\rm J,ion}]$ and $M_{\rm halo}>M_{\rm J,ion}$. The former, together with molecular cooling haloes, is not expected to form stars after reionization. Therefore, their contributions to the Pop~III SFRD are removed for the Pop~III SFRD models in the next subsection. To evaluate the relative importance of the three groups, we plot the halo mass distribution of the representative sample in Fig.~\ref{mhdis}, where haloes are weighted by enclosed mass of active Pop~III stars, $M_{\rm PopIII}$, such that the distribution is proportional to $dM_{\rm PopIII}/d\log M_{\rm halo}$. It turns out that the ratio of the contributions from the three groups to Pop~III star formation is approximately 2 : 1 : 1. Besides, the distribution at $M_{\rm halo}\gtrsim M_{\rm J,ion}$ can be approximated with a power-law of index $\alpha_{m}\sim 0.5$ (solid), while that at $M_{\rm th}^{\rm atom}\lesssim M_{\rm halo}\lesssim M_{\rm J,ion}$ can be described by another power-law with $\alpha_{m}\sim -1$.

\begin{figure}
\includegraphics[width=1\columnwidth]{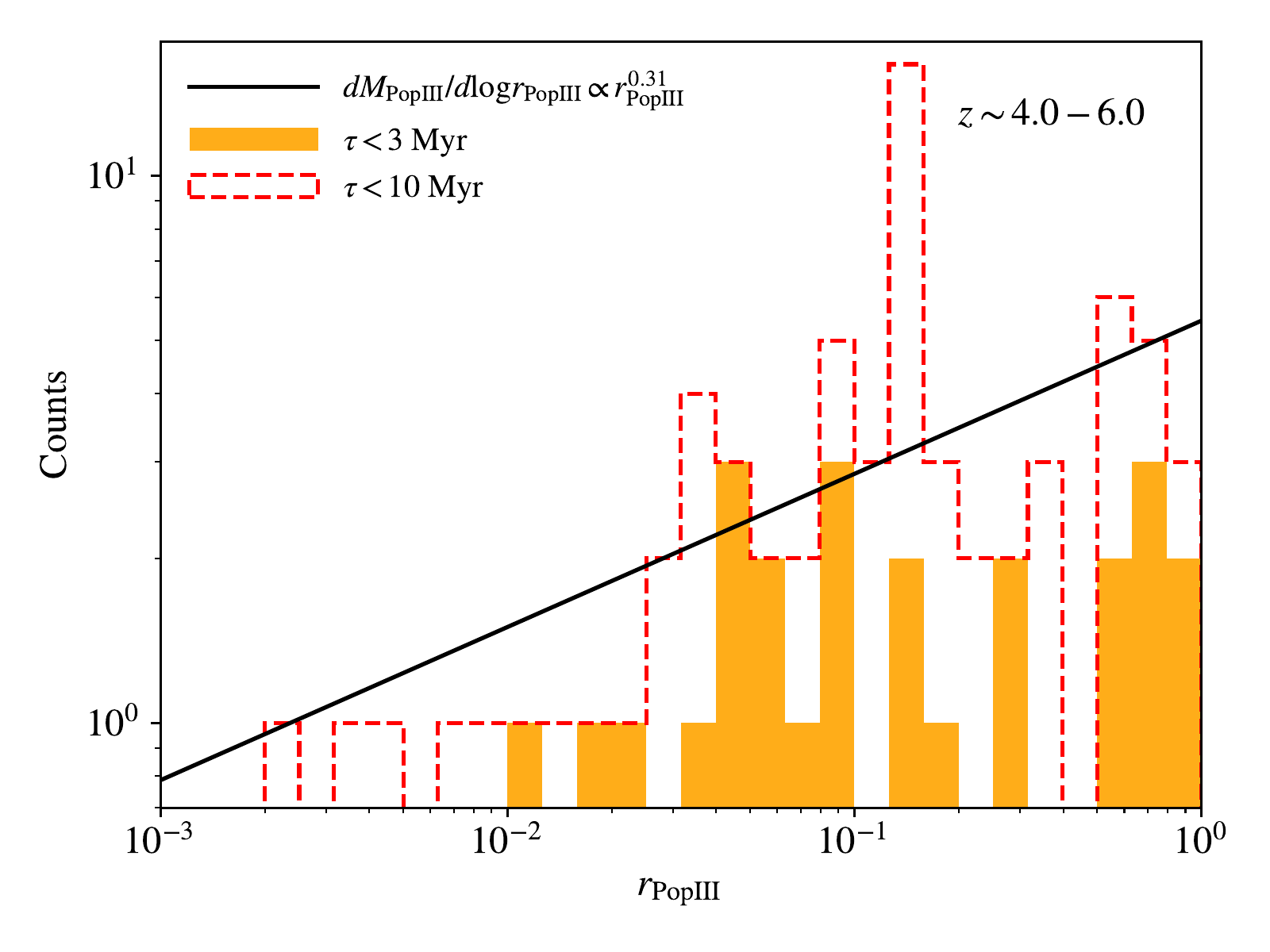}
\vspace{-20pt}
\caption{Distribution of relative distances $r_{\rm PopIII}\equiv R_{\rm PopIII}/R_{\rm vir}$ (in log scale) for the active Pop~III particles with ages $\tau<3$ (orange histograms) and 10 (red dashed contour) Myr, in haloes above the Jeans mass of ionized gas ($M_{\rm halo}>M_{\rm J,ion}$) from the representative sample. The distribution is fitted to a power-law form, resulting in a power-law index of $\alpha_{r}\sim 0.3$, such that the enclosed mass of active Pop~III stars follows $M_{\rm PopIII}(<r)\propto r^{\alpha_{r}}\sim r^{0.3}$. }
\label{rratdis}
\end{figure}

Besides the host mass, another crucial property of late-time Pop~III star formation is the distribution of Pop~III stars in their host haloes. We define the relative distance, $r_{\rm PopIII}$, from an active Pop~III particle to the halo center as the ratio of the physical distance $R_{\rm PopIII}$ to the virial radius $R_{\rm vir}$, i.e. $r_{\rm PopIII}\equiv R_{\rm PopIII}/R_{\rm vir}$. The distribution of $r_{\rm PopIII}$ is shown in Fig.~\ref{rratdis} for the active Pop~III particles in haloes with $M_{\rm halo}>M_{\rm J,ion}$ from the representative sample. This distribution, i.e. $dM_{\rm PopIII}/d\log r_{\rm PopIII}\propto r_{\rm PopIII}^{3}\rho_{\rm PopIII}(r_{\rm PopIII})$, given the density profile of Pop~III stars, $\rho_{\rm PopIII}$, can be approximated with a power-law of index $\alpha_{r}\sim 0.3$. The result for all atomic cooling haloes ($M_{\rm halo}>M_{\rm th}^{\rm atom}$) is similar. This indicates that the (quasi-natal) distribution of Pop~III stars is less concentrated than that of Pop~II/I stars and dark matter (with $\rho\propto r^{-4}$ and $r^{-3}$, i.e. $\alpha_{r}\sim -1-0$, at the outskirts). About half (47-61\%) of the Pop~III particles occur at the outskirts of haloes ($r_{\rm PopIII}\gtrsim 0.1$), consistent with the `Pop~III wave' theory \citep{tornatore2007population}. 
However, a few percent of Pop~III particles with ages $\tau\sim 3-10$~Myr are still found in halo centers ($r_{\rm PopIII}\lesssim 10^{-2}$), which are expected to be polluted by metals. One explanation is that for haloes during assembly (mergers), the mass center of a halo as a whole may not be close to any sub-haloes with recent star formation activities (i.e. sources of metal enrichment), as shown in Fig.~\ref{gasdis}. Nevertheless, outflows driven by SN winds may have enriched the halo center (or even the entire halo) in reality, so that we may have overestimated Pop~III star formation. 

\begin{figure}
\includegraphics[width=1\columnwidth]{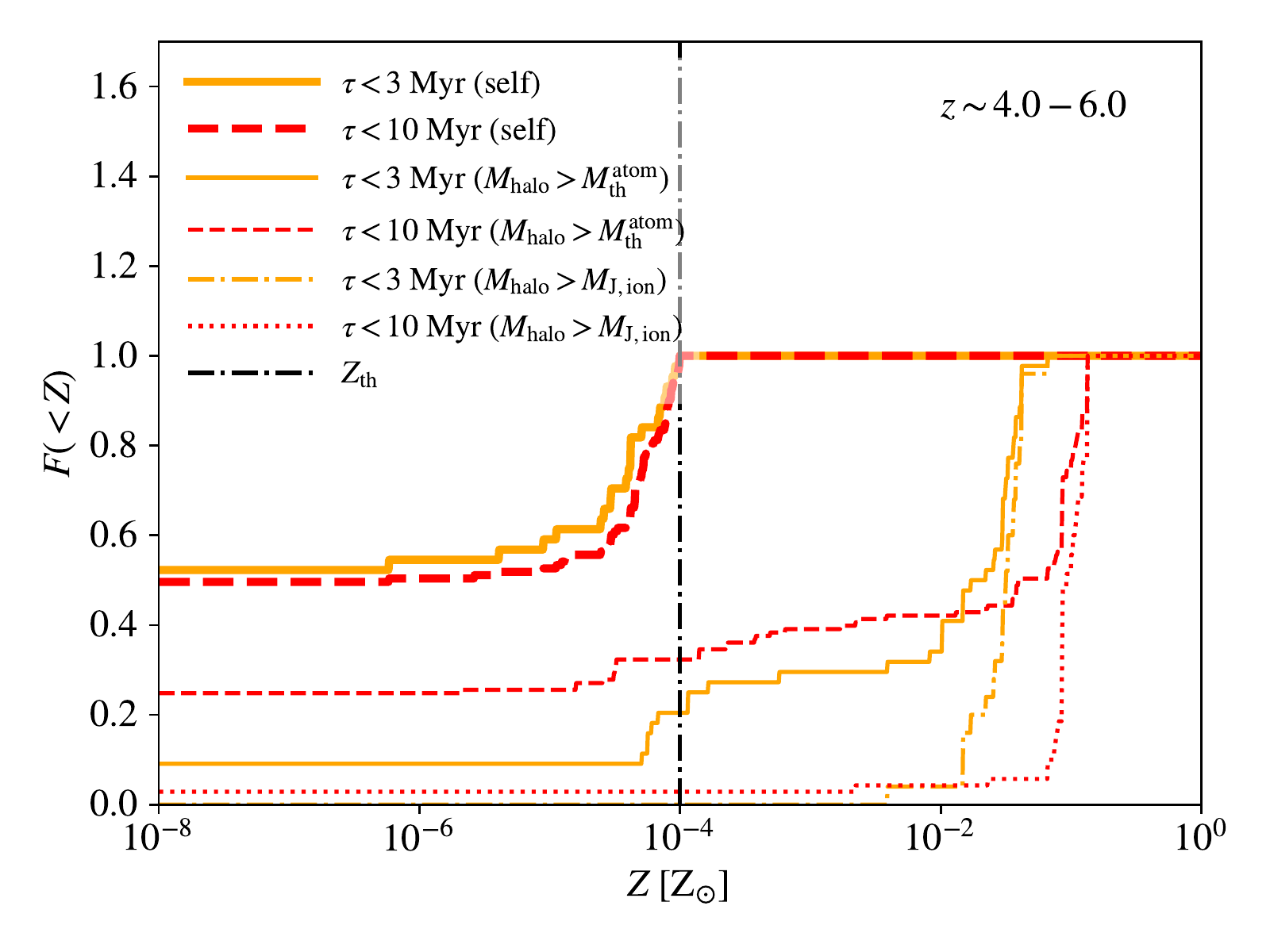}
\vspace{-20pt}
\caption{Cumulative metallicity distribution functions for active Pop~III particles (thick) and their host haloes (thin), from the representative sample. The results for atomic cooling haloes are shown with solid and dashed curves, while those for haloes above the Jeans mass of ionized gas with dashed-dotted and dotted curves, for $\tau<3$ and 10 Myr, respectively. Again, haloes are weighted by enclosed masses of active Pop~III stars. To better capture the natal environments of Pop~III stars, gas-phase metallicity is adopted for host haloes with $\tau<3$ Myr, while stellar metallicity is used for $\tau<10$~Myr. }
\label{cumZ}
\end{figure}

In light of this, we further look into the extreme case in which metals are fully mixed in the entire halo (i.e. within $R_{\rm vir}$) by measuring the mean (gas-phase and stellar) metallicities of Pop~III host haloes in the representative sample. The cumulative distribution functions of the halo mean metallicities for different groups of haloes are shown in Fig.~\ref{cumZ}, together with the metallicities of active Pop~III particles themselves. The latter is meant to explore the dependence of Pop~III star formation on the critical metallicity ($Z_{\rm crit}\sim 10^{-6}-10^{-3.5}\ \rm Z_{\odot}$) for the Pop~III to Pop~II/I transition. Atomic cooling haloes ($M_{\rm halo}>M_{\rm th}^{\rm atom}$, representative before reionization) and haloes above the filtering mass ($M_{\rm halo}>M_{\rm J,ion}$, representative after reionization) are considered separately. To better capture the natal environments of Pop~III stars, we focus on the gas-phase metallicity for haloes\footnote{For a Pop~III particle, $\tau$ is the age of the underlying stellar population. For a halo with recent Pop~III star formation, $\tau$ is the age of the youngest Pop~III stellar particles within it.} with $\tau<3$ Myr, but stellar metallicity for haloes with $\tau<10$~Myr. It is shown that if the Pop~III mode is restricted to metal-free gas (equivalent to $Z_{\rm crit}\lesssim 10^{-6}\ \rm Z_{\odot}$ in our case), about 50\% of Pop~III star formation will be shifted to Pop~II/I. If metals are fully mixed inside haloes and $Z_{\rm crit}\lesssim 10^{-5}\ \rm Z_{\odot}$, $\sim 10-25$\% of Pop~III star formation remains before reionization (for $M_{\rm halo}>M_{\rm th}^{\rm atom}$), while only $\lesssim 3$\% remains after reionization (for $M_{\rm halo}>M_{\rm J,ion}$).

\begin{figure}
\includegraphics[width=1\columnwidth]{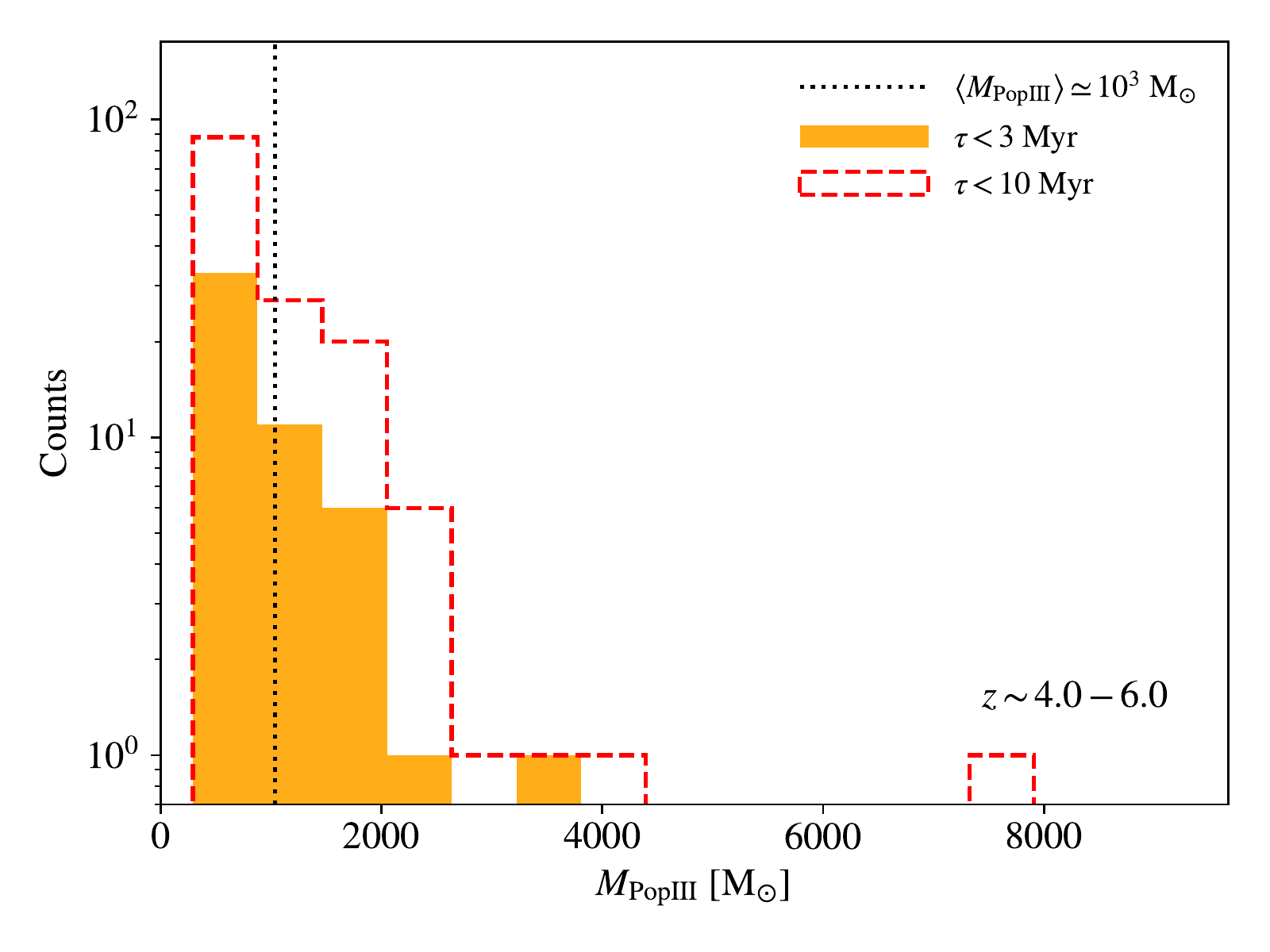}
\vspace{-20pt}
\caption{Distribution of enclosed mass of active Pop~III stars for the representative sample of recent Pop~III star formation within 3 (orange histograms) and 10 (red dashed contour) Myr. There is no clear correlation between $M_{\rm PopIII}$ and $M_{\rm halo}$, such that the distribution remains similar when only haloes with $M_{\rm halo}>M_{\rm J, ion}$ are considered (i.e. after reionization). 
The average mass of active Pop~III stars per halo $\langle M_{\rm PopIII}\rangle \simeq 10^{3}\ \rm M_{\odot}$ is shown with the vertical dotted line.} 
\label{MpopIII}
\end{figure}

Finally, a parameter of particular importance for direct detection of Pop~III stars is the total mass of active Pop~III stars $M_{\rm PopIII}$ per halo. This parameter is the product of the `quantum' of Pop~III star formation, i.e. the typical Pop~III stellar mass formed per local (cloud-scale) star formation event, and the number of Pop~III star-forming clouds coexisting in a few Myr. 
Fig.~\ref{MpopIII} shows the distribution of $M_{\rm PopIII}$ for the entire representative sample. We find no clear correlation between $M_{\rm PopIII}$ and $M_{\rm halo}$, such that the distribution remains similar when only haloes with $M_{\rm halo}>M_{\rm J, ion}$ are considered (i.e. after reionization). The average mass of active Pop~III stars per halo is $\langle M_{\rm PopIII}\rangle \simeq 10^{3}\ \rm M_{\odot}$ with large scatter. More than 50\% of haloes only have one active Pop~III particle (i.e. $M_{\rm PopIII}=m_{\star}\simeq 600\ \rm M_{\odot}$), and less than 10\% of haloes have $M_{\rm PopIII}\sim 2\times 10^{3}-10^{4}\ \rm M_{\odot}$, consistent with theoretical and observational upper limits of $M_{\rm PopIII}\lesssim 10^{6}\ \rm M_{\odot}$ \citep{yajima2017upper,bhatawdekar2020}\footnote{\citet{yajima2017upper} derived $M_{\rm PopIII}\lesssim 10^{6}\ \rm M_{\odot}$ from a semi-analytical model for the collapse of primordial gas under the effect of angular momentum loss via Lyman-$\alpha$ (Ly$\alpha$) radiation drag and the gas accretion onto a galactic centre. The lack of evidence for Pop~III dominated systems in the Hubble Frontier Fields at $z\sim 6-9$ \citep{bhatawdekar2020} implies $M_{\rm PopIII}\lesssim 4-7\times 10^{5}\ \rm M_{\odot}$, given a limiting rest-frame UV (absolute) AB magnitude $M_{\rm UV}=-13.5$ (see Sec.~\ref{s4.3} for the Pop~III stellar population synthesis model adopted to derive $M_{\rm PopIII}$ from $M_{\rm UV}$).}. Our results also (marginally) agree with a recently discovered strongly lensed Pop~III candidate Lyman-$\alpha$ (Ly$\alpha$) emitter at $z\simeq 6.6$, which has $M_{\rm PopIII}\sim 10^{4}\ \rm M_{\odot}$ \citep{vanzella2020candidate}, residing at the high mass end of our prediction. 
Note that the simulations of \citet{xu2016late,danielle2020} also find typically $M_{\rm PopIII}\lesssim 10^{3}\ \rm M_{\odot}$, while other simulations with lower resolution or different star formation routines predict higher values, e.g. $M_{\rm PopIII}\gtrsim 10^{5}\ \rm M_{\odot}$ \citep{pallottini2014simulating,sarmento2018following}. As the observational constraints are still weak/unclear, the total mass of active Pop III stars per halo/galaxy is uncertain, especially for massive haloes at late times ($M_{\rm halo}\gtrsim 10^{9}\ \rm M_{\odot}$, $z\lesssim 6$), depending on resolution and sub-grid models for star formation and stellar feedback, particularly chemical feedback from SNe.

\subsection{Extrapolating Pop~III star formation to the present day}
\label{s4.2}
Based on what is learned from the representative sample, we now extrapolate Pop~III star formation to $z=0$ by introducing corrections to the fit of simulated Pop~III SFRD $\dot{\rho}_{\star,\rm PopIII}^{\rm sim}$ (Equ.~\ref{fit}) for reionization and metal mixing. We decompose the Pop~III SFRD into two components: one from molecular cooling haloes ($M_{\rm halo}<M_{\rm th}^{\rm atom}$) and the other from atomic cooling haloes ($M_{\rm halo}>M_{\rm th}^{\rm atom}$). Both components are subject to reionization corrections, while we only consider additional metal mixing for the latter. Actually, in our simulation, most molecular cooling haloes only experience one episode of Pop~III star formation before merging into more massive haloes\footnote{In the representative sample, $\sim 80$\% of molecular cooling haloes hosting active Pop~III stars ($\tau<3$~Myr) have not experienced any previous star formation (and internal enrichment).}, such that internal enrichment is not important. 
Although external enrichment may play a role \citep{wise2014birth,smith2015first,jeon2017}, we neglect this effect for simplicity.  Note that star formation in molecular cooling haloes is prohibited after reionization, and the contribution of molecular cooling haloes is only a few percent during reionization ($z\sim 7-13$) due to strong LW feedback (see Fig.~\ref{fiso}). We write the Pop~III SFRD after such corrections as
\begin{align}
\dot{\rho}_{\star,\rm PopIII}^{\rm cor}=\dot{\rho}_{\star,\rm PopIII}^{\rm sim}(\hat{f}_{\rm atom}\langle f_{\rm mp}\rangle+\hat{f}_{\rm mol})\ ,\label{e1}
\end{align}
where $\hat{f}_{\rm mol}$ and $\hat{f}_{\rm atom}$ are the terms for reionization correction, while $\langle f_{\rm mp}\rangle$ captures the effect of additional metal mixing. The reionization terms are calculated with
\begin{align}
    &\hat{f}_{k}(z)=f_{k}(z)\times \{f_{k,0}+f_{k,1}[1-\hat{f}_{\rm ion}(z)]\}\ ,\label{e9}
\end{align}
where $f_{k}(z)$ is derived from the simulation for $k=\rm mol,\ atom$, with $f_{\rm mol}+f_{\rm atom}=1$. To be specific, we use a smoothed version of $f_{\rm mol}$ based on simulation data (see Fig.~\ref{fiso} and the left panel of Fig.~\ref{radbg}), in which $f_{\rm mol}=1$ for $z>19$ with negligible LW feedback, $f_{\rm mol}=0.05$ for $12.5>z>6$ under a strong LW background ($J_{\rm LW,bg,21}\gtrsim 1$), $f_{\rm mol}=0.2$ for $z<6$ under a moderate LW background ($J_{\rm LW,bg,21}\sim 0.1-1$), and these three plateaus are connected with two linear functions of $z$. 
Within each component $k$, $f_{k,0}$ is the fraction of Pop~III star formation unaffected by reionization (i.e. in massive haloes $M_{\rm halo}>M_{\rm J,ion}$), and $f_{k,1}$ is that suppressed by reionization. Note that $f_{k,0}+f_{k,1}=1$. We set $f_{\rm mol,0}=0$ and $f_{\rm atom,0}=0.5$, based on the representative sample (see Fig.~\ref{mhdis}).

For (additional) metal mixing, $\langle f_{\rm mp}\rangle$ is defined as the fraction of Pop~III star formation remaining, when more sufficient metal mixing is considered than captured in our simulation:
\begin{align}
&\langle f_{\rm mp}\rangle=\int_{M_{1}}^{M_{2}}f_{\rm mp}(z, M)w(M)dM/\int_{M_{1}}^{M_{2}}w(M)dM\ ,\label{e2}\\
&f_{\rm mp}(z,M)=\max\{1-\left[R_{\rm mix}/R_{\rm vir}\right]^{\alpha_{r}}, 0\}\ ,\label{e3}
\end{align}
where $f_{\rm mp}(z,M)$ is the metal-poor fraction of potential Pop~III forming gas as a function of halo virial radius $R_{\rm vir}$ and metal mixing radius $R_{\rm mix}$, for a halo of mass $M$ at $z$. In the second line (Equ.~\ref{e3}), we have assumed spherical symmetry and locate the halo center as the source of enrichment. We adopt $M_{1}=M_{\rm th}^{\rm atom}$ and $M_{2}=\max[10M_{\rm th}^{\rm atom},M_{\rm crit}(\nu=2)]$, corresponding to the mass range of haloes in our simulation, where $M_{\rm crit}(\nu=2)$ is the critical mass for 2-sigma peaks. The weight function is written as $w(M_{\rm halo}) = A M_{\rm halo}^{\alpha_{m}-1}\propto M_{\rm halo}^{-1}dM_{\rm PopIII}/d\log M_{\rm halo}$, where $\alpha_{m}\sim 0.5$ and $A=0.4$ for $M_{\rm halo}\ge M_{\rm J,ion}$, while $\alpha_{m}\sim-1$ and $A=1-\hat{f}_{\rm ion}(z)$ for $M_{\rm halo}<M_{\rm J,ion}$. The power-law indices $\alpha_{m}$ and normalization factors $A$ are derived from the simulated distribution of (active) Pop~III mass in haloes, $dM_{\rm PopIII}/d\log M_{\rm halo}$, as shown in Fig.~\ref{mhdis}. Note that the reionization effect has been absorbed into $A$ for low-mass haloes ($M_{\rm halo}<M_{\rm J,ion}$). We use $\alpha_{r}\sim 0.3$, based on the radius distribution of Pop~III particles (Fig.~\ref{rratdis}). The metal mixing radius $R_{\rm mix}$ is estimated by tracking the halo growth history with the gravity-driven turbulent diffusion model based on \citet{karlsson2008uncovering},  
\begin{align}
R_{\rm mix}(z, M)&=\left[6\int_{z_{i}}^{z}D_{\rm turb}(z')\left |\frac{dt}{dz'}\right|dz'\right]^{1/2}\ ,\label{e4}\\
D_{\rm turb}(z')&\equiv\langle v_{\rm turb}\rangle l_{\rm turb}/3=\beta_{\rm mix}v'_{\rm vir}R'_{\rm vir}/3\ ,\label{e5}\\
v'_{\rm vir}&=\sqrt{\frac{GM'}{R'_{\rm vir}}}\ ,\quad R'_{\rm vir}=\left[\frac{3 M'}{4\pi \Delta\rho_{m}(z')}\right]^{1/3}\ .\label{e6}
\end{align}
Here $\Delta=200$, and $\beta_{\rm mix}$ is an adjustable parameter that reflects the strength of metal mixing, in terms of how the turbulent diffusion coefficient $D_{\rm turb}(z')\equiv D_{\rm turb}(z'|z,M)$ depends on halo dynamics. Evidently, $f_{\rm mp}(z,M)$ decreases with increasing $\beta_{\rm mix}$. The onset of internal metal enrichment $z_{i}\equiv z_{i}(z,M)$ is derived by $M'(z_{i}|z, M)=M_{\rm th}^{\rm atom}(z_{i})$ for $z_{i}<20$. The halo growth history is obtained by solving for $M'\equiv M'(z'|z, M)$, which is the progenitor mass at $z'>z$ of a halo at $z$ with mass $M$. This is done by integrating the (average) halo growth rate formula from \citet{fakhouri2010merger}, which is derived from simulations for $\Lambda$CDM cosmology:
\begin{align}
    \frac{dM}{dz}&=\dot{M}(z,M)\left |\frac{dt}{dz}\right|\ ,\notag\\
    &\simeq 46\ \mathrm{M_{\odot}\ yr^{-1}}\left(\frac{M}{10^{12}\ \mathrm{M_{\odot}}}\right)^{1.1}\xi(z)\left |\frac{dt}{dz}\right|\ ,\notag\\
    \xi(z)&=\left[1.1(1+z)-0.11\right]\sqrt{\Omega_{m}(1+z)^{3}+\Omega_{\Lambda}}\ .\label{e7}
\end{align}
In general, our model predicts that $f_{\rm mp}(z,M)$ increases with increasing mass $M$ and increasing redshift $z$.

We also consider a more simulation-based model, where the metal-poor fraction is expressed with
\begin{align}
    \hat{f}_{\rm mp}^{\rm sim}(z) = f_{\rm mp}^{\rm post} + (f_{\rm mp}^{\rm pre}-f_{\rm mp}^{\rm post})[1-\hat{f}_{\rm ion}(z)]\ .\label{e8}
\end{align}
Here $f_{\rm mp}^{\rm post}$ and $f_{\rm mp}^{\rm pre}$ are the fractions of metal-poor gas for Pop~III star formation in atomic cooling haloes after and and before reionization. If metal mixing is actually efficient at the halo scale ($R_{\rm mix}\gtrsim R_{\rm vir}$), but not fully captured in our simulation, Pop~III particles in haloes with (mass-weighted) \textit{mean} metallicities above $Z_{\rm crit}$ should be removed. This pessimistic case can be evaluated with the distributions of halo mean metallicities for Pop~III host haloes, as shown in Fig.~\ref{cumZ} for the representative sample. As an upper limit, we use $f_{\rm mp}^{\rm pre}\simeq 0.25$ and $f_{\rm mp}^{\rm post}\simeq 0.03$, given a critical metallicity $Z_{\rm crit}\lesssim 3\times 10^{-5}\ \rm Z_{\odot}$, based on the halo stellar metallicity distributions of Pop~III particles with ages $\tau<10$~Myr, in all atomic cooling haloes ($M_{\rm halo}>M_{\rm th}^{\rm atom}$, before reionization) and only massive haloes above the Jeans mass of ionized gas ($M_{\rm halo}>M_{\rm J,ion}$, after reionization). 

\begin{figure}
\centering
\includegraphics[width=1\columnwidth]{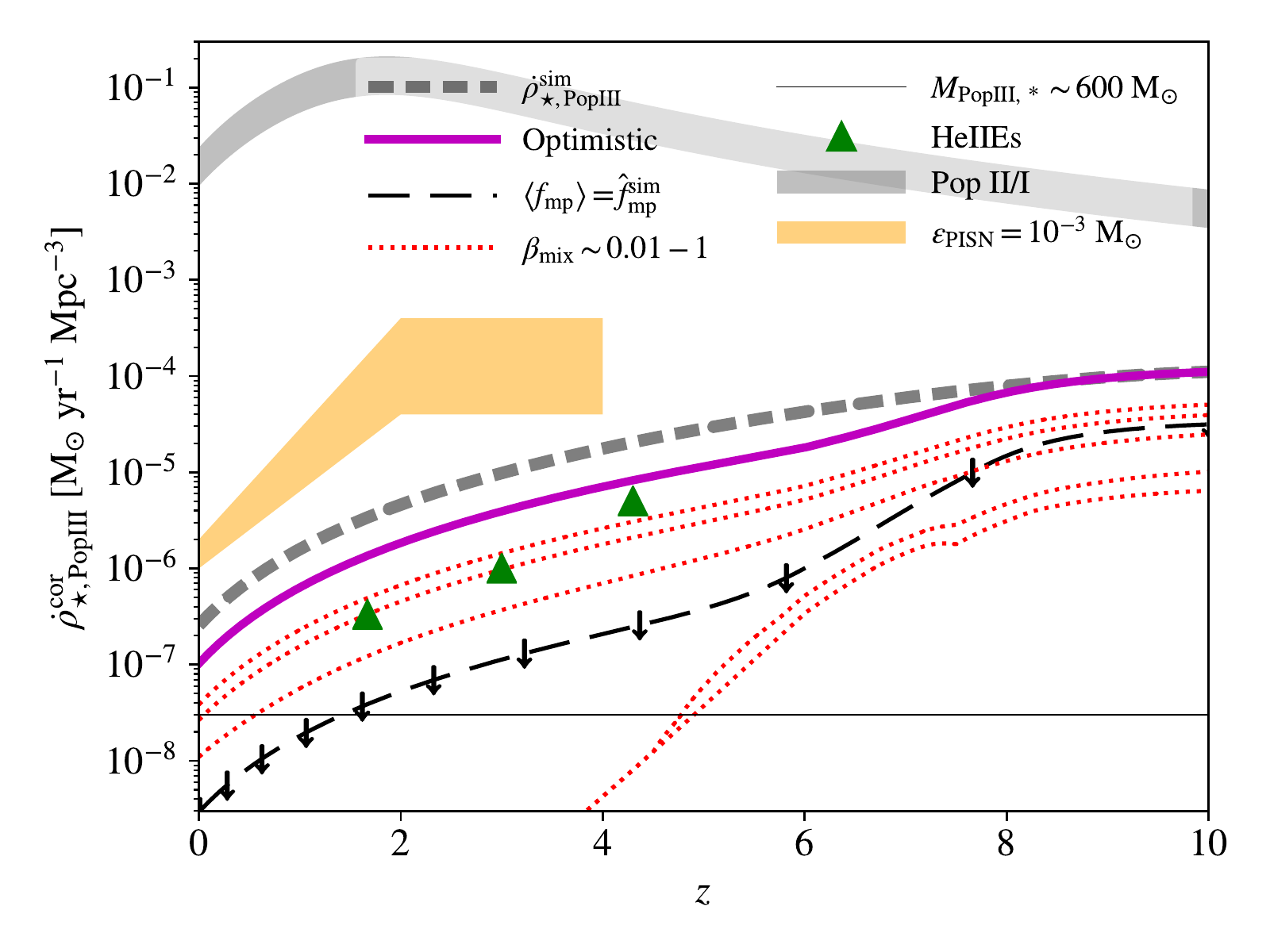}
\vspace{-20pt}
\caption{Pop~III SFRD models based on extrapolation of the simulation results (thick short dashed), with corrections for reionization (Equ.~\ref{e9}) and different models of metal mixing. The optimistic model ($\langle f_{\rm mp}\rangle=1$) is plotted with the thick solid curve. The pessimistic model based on the metal-poor fraction $\hat{f}_{\rm mp}^{\rm sim}$ with halo-scale metal mixing of \textit{simulated} haloes (Equ.~\ref{e8}) is shown with the (normal) long dashed curve of downward arrows. Semi-analytical metal mixing models (Equ.~\ref{e2}-\ref{e7}) for $\beta_{\rm mix}=0.01$, 0.03, 0.1, 0.3 and 1 are shown with the dotted curves (from top to bottom). The thin horizontal line shows the physically-motivated `critical' Pop~III SFRD for a typical halo ($M_{\rm halo,*}\sim 2\times 10^{12}\ \rm M_{\odot}$) at $z=0$ to form one typical Pop~III star cluster of $\sim 600\ \rm M_{\odot}$ within one dynamical timescale $t_{\rm dyn}\sim 0.1/H_{0}$ (see Sec.~\ref{s4.3} for details). 
The orange shaded region shows the upper bounds inferred from observations of super-luminous SNe (see main text). Constraints from narrow \HeII\ line emitters (HeIIEs) as candidates of Pop~III systems \citep{nagao2008photometric,prescott2009discovery,cassata2013he} are shown with the triangles. The Pop~II/I SFRD from \citet{madau2014araa} is also shown for comparison (grey shaded region). }
\label{sfrdex}
\end{figure}

Finally, examples of the Pop~III SFRD models with the above reionization and metal-mixing corrections, $\dot{\rho}_{\star,\rm PopIII}^{\rm cor}$, are shown in Fig.~\ref{sfrdex}, on top of observational constraints and the Pop~II/I counterpart \citep{madau2014araa}. We consider the upper bounds on Pop~III SFRD ($\sim 10^{-6}\ \rm M_{\odot}\ yr^{-1}\ Mpc^{-3}$ at $z\sim 0$ and $\sim 10^{-4}\ \rm M_{\odot}\ yr^{-1}\ Mpc^{-3}$ at $z\sim 2-4$), inferred from the observed rate densities of super-luminous SNe (as PISN candidates, see \citealt{gal2012luminous,cooke2012superluminous}), assuming a Pop~III PISN efficiency of $\epsilon_{\rm PISN}=10^{-3}\ \rm M_{\odot}$ (for typical top-heavy IMFs). Our model SFRDs are always lower than these upper bounds (by at least a factor of 10), even for the optimistic case with $\langle f_{\rm mp}\rangle=1$. We also plot the Pop~III SFRD values, inferred from observations of narrow \HeII\ line emitters as candidates of Pop~III systems \citep{nagao2008photometric,prescott2009discovery,cassata2013he}, which are generally lower (by up to a factor of 4) than the optimistic model ($\langle f_{\rm mp}\rangle=1$) and approximately correspond to the metal mixing models with $\beta_{\rm mix}\lesssim 0.03$. Note that such observational constraints are highly sensitive to the Pop~III IMF, escape fraction of ionizing photons and potential selection effects.

\subsection{Termination of Pop~III star formation}
\label{s4.3}

The precise definition of what it means to terminate Pop~III star formation is non-trivial, in the absence of abrupt cut-offs in the Pop~III SFRD, akin to a cosmic phase transition such as reionization. Our theoretical models here indeed do not exhibit any precipitous drop, and are instead characterized by a more gradual tapering off. We first consider a physically motivated definition based on the average Pop~III star formation rate\footnote{With the SFR formula~(Equ.~\ref{mdot}), our models (see Fig.~\ref{sfrdex}) can predict the \textit{average} formation rate of Pop~III stars in a present-day halo of a mass similar to that of the Milky Way halo ($M_{\rm halo}\simeq 1.5\times 10^{12}\ \rm M_{\odot}$). The result for the optimistic model ($\langle f_{\rm mp}\rangle=1$) is $\dot{M}_{\rm PopIII}\sim 10^{-6}\ \rm M_{\odot}\ yr^{-1}$. Given $M_{\rm PopIII}\sim 10^{3}\ \rm M_{\odot}$, this \textit{average} SFR implies that the probability of such a halo to have recent Pop~III star formation (within 3~Myr) is $\sim 0.003$, which serves as a rough estimation for the Milky Way. More accurate estimation for the chance of finding active Pop~III stars in the real Milky Way halo needs to consider the detailed assembly history and metal mixing process.} 
(SFR) for a halo of mass $M$ at $z$, given a Pop~III SFRD $\dot{\rho}_{\rm \star,PopIII}$:
\begin{align}
    \dot{M}_{\rm PopIII}&(M,z|\dot{\rho}_{\rm \star,PopIII})=V_{\rm eff}(M,z)\dot{\rho}_{\star,\rm PopIII}\ ,\notag\\
    V_{\rm eff}&(M,z) = w(M)\left[n_{\rm h}(M,z)\int_{M_{1}}^{M_{2}}w(M)dM\right]^{-1}\ ,\label{mdot}
\end{align}
where $n_{\rm h}$ is the halo mass function (calculated by \citealt{murray2013hmfcalc}), $w(M)$, $M_{1}$ and $M_{2}$ refer to the weight function and mass range of Pop~III hosts (see Equ.~\ref{e2} and the description thereafter), which embodies the distribution of Pop~III mass in haloes, i.e. $w(M_{\rm halo})\propto M_{\rm halo}^{-1}dM_{\rm PopIII}/d\log M_{\rm halo}$ (see Fig.~\ref{mhdis}).

We then focus on a typical halo at $z=0$ with a mass $M_{\rm halo,*}\sim 2\times 10^{12}\ \rm M_{\odot}$ and define the `critical' Pop~III SFRD as the one required to form one typical/minimum Pop~III star cluster with $M_{\rm PopIII,*}\sim 600\ \rm M_{\odot}$ in one dynamical timescale $t_{\rm dyn}\sim 0.1/H_{0}$. This leads to $\dot{\rho}_{\star, \rm crit}\equiv M_{\rm PopIII,*}/[t_{\rm dyn}V_{\rm eff}(M=M_{\rm halo,*},z=0)]\sim 3\times 10^{-8}\ \rm M_{\odot} \rm\, yr^{-1} \rm\, Mpc^{-3}$. Finally, the termination of Pop~III star formation is defined via $\dot{\rho}_{\star,\rm PopIII}=\dot{\rho}_{\star,\rm crit}$. In our case, Pop~III star formation will be terminated at $z\sim 5$ with complete halo-scale metal mixing (i.e. $\langle f_{\rm mp}\rangle\sim 0$, achieved with $\beta_{\rm mix}\gtrsim 0.18$). For the simulation-based pessimistic model (i.e. $\langle f_{\rm mp}\rangle=\hat{f}_{\rm mp}^{\rm sim}$, see Equ.~\ref{e8}), Pop~III star formation ends at $z\gtrsim 1.5$, which approximately corresponds to the case of $\beta_{\rm mix}\gtrsim 0.15$. While for inefficient metal mixing with $\beta_{\rm mix}\lesssim 0.03$, there is no termination at $z>0$, according to this definition.

\begin{figure*}
\hspace{-5pt}
\centering
\subfloat[Pop~III stars, $ M_{\rm PopIII} =10^{3}\ \rm M_{\odot}$]{\includegraphics[width= 1\columnwidth]{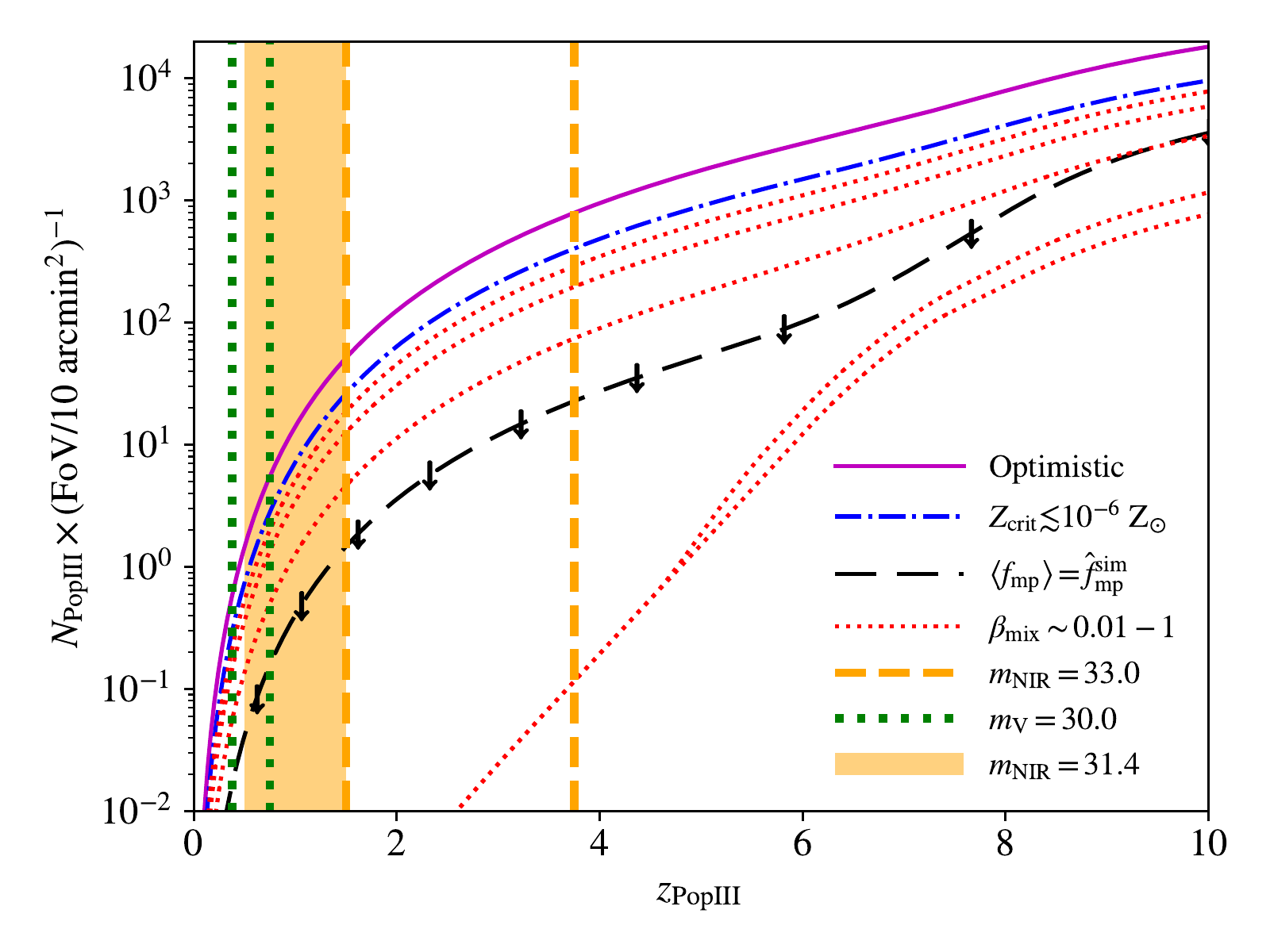}}
\subfloat[Pop~III PISNe, $\epsilon_{\rm PISN}=10^{-3}\ \rm M_{\odot}^{-1}$]{\includegraphics[width= 1\columnwidth]{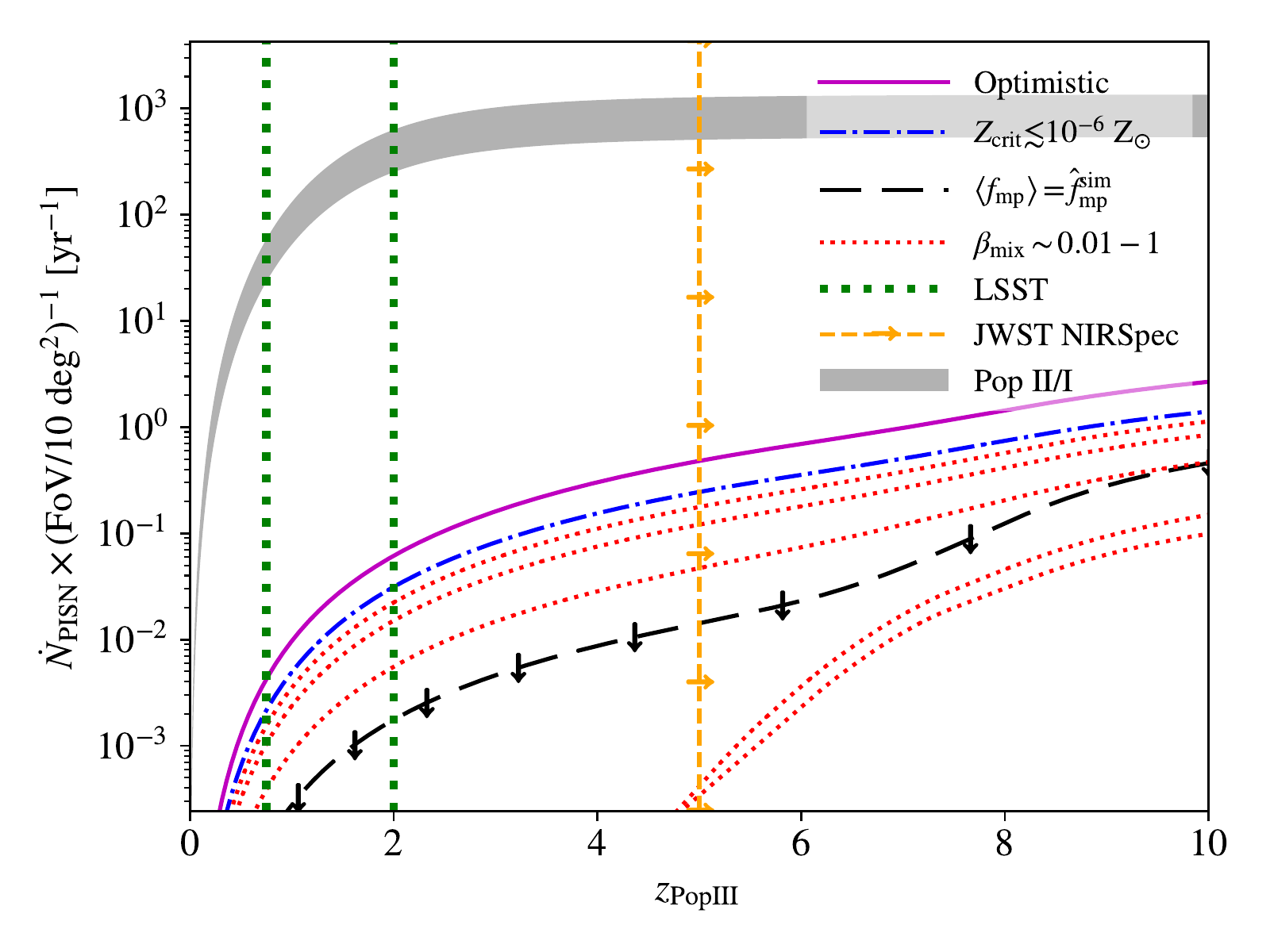}}
\vspace{-10pt}
\caption{Detectablility of Pop~III systems, for the optimistic model (solid), the strictly metal-free case with $Z_{\rm crit}\lesssim 10^{-6}\ \rm Z_{\odot}$ (dashed-dotted), the pessimistic model based on the metal-poor fraction $\hat{f}_{\rm mp}^{\rm sim}$ with halo-scale metal mixing of \textit{simulated} haloes (long-dashed), and semi-analytical metal mixing models with $\beta_{\rm mix}=0.01$, 0.03, 0.1, 0.3 and 1 (dotted, from top to bottom). The underlying Pop~III SFRD models are shown in Fig.~\ref{sfrdex}. Left panel (a): Number of Pop~III host systems per $10\ \rm arcmin^{2}$ as a function of horizon redshift $z_{\rm PopIII}$, assuming that all Pop~III stars are grouped into systems with $ M_{\rm PopIII} =10^{3}\ \rm M_{\odot}$. The horizon redshifts for the JWST NIRCam filter F150W are shown with the shaded region ($z_{\rm PopIII}\sim 0.5-1.5$) and thick dashed vertical lines ($z_{\rm PopIII}\sim 1.5-3.75$), given the limiting (AB) magnitudes 31.4 and 33 for ultra-deep campaigns and lensing, respectively. Similarly, for the HST WFC3 filter F555W, we have $z_{\rm PopIII}\sim 0.38-0.75$ (thick dotted), given a limiting magnitude of $m_{\rm V}=30$ (for $\rm SNR>5$ in a 10-hour exposure). 
In each case, we derive the magnitudes of Pop~III stars of $M_{\rm PopIII}=1000-5000\ \rm M_{\odot}$ from the SPS code \textsc{yggdrasil} \citep{zackrisson2011spectral}, under their Pop~III.1 model (see the text of Sec.~\ref{s4.3} for details). 
Right panel (b): PISN detection rates per $10\ \rm deg^{2}$ as a function of horizon redshift $z_{\rm PopIII}$, assuming a typical PISN efficiency $\epsilon_{\rm PISN}=10^{-3}\ \rm M_{\odot}^{-1}$ for Pop~III. The detection limits for LSST (g, r, i and z bands) with and without circumstellar medium interactions are shown with the thick vertical dotted lines ($z_{\rm PopIII}\sim 0.75-2$), while that for JWST NIRSpec ($\rm SNR>10$) is shown with the orange dashed line ($z_{\rm PopIII}\sim 5$), based on the properties of PISN candidate SN2016aps \citep{nicholl2020}. We also plot an upper limit of the PISN rate for Pop~II/I stars with the gray shaded region, assuming no mass loss by stellar winds (see main text for details).}
\label{Drate}
\end{figure*}

\begin{figure*}
\hspace{-5pt}
\centering
\subfloat[Occupation fraction]{\includegraphics[width= 1\columnwidth]{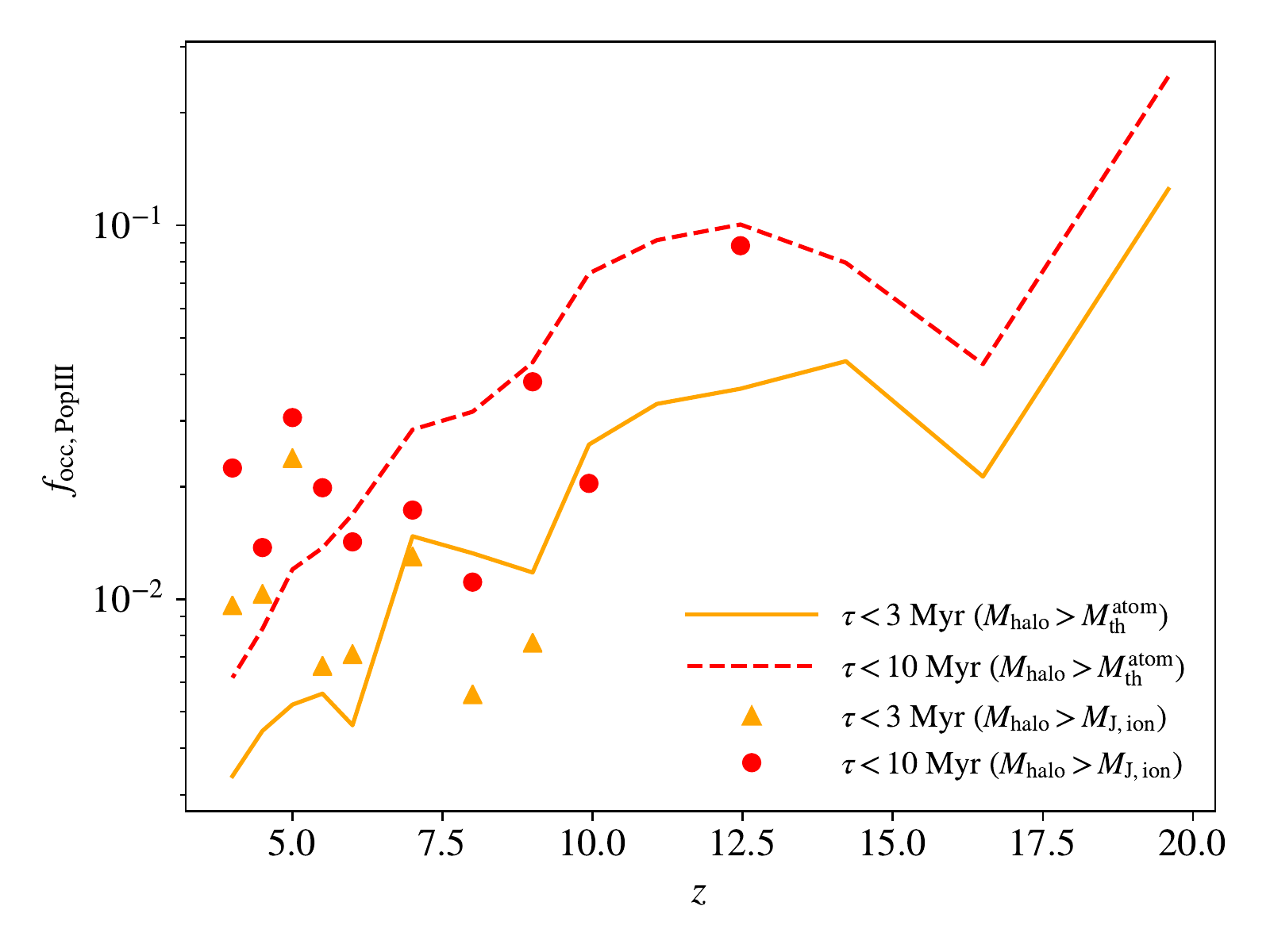}}
\subfloat[Mass ratio distribution]{\includegraphics[width= 1\columnwidth]{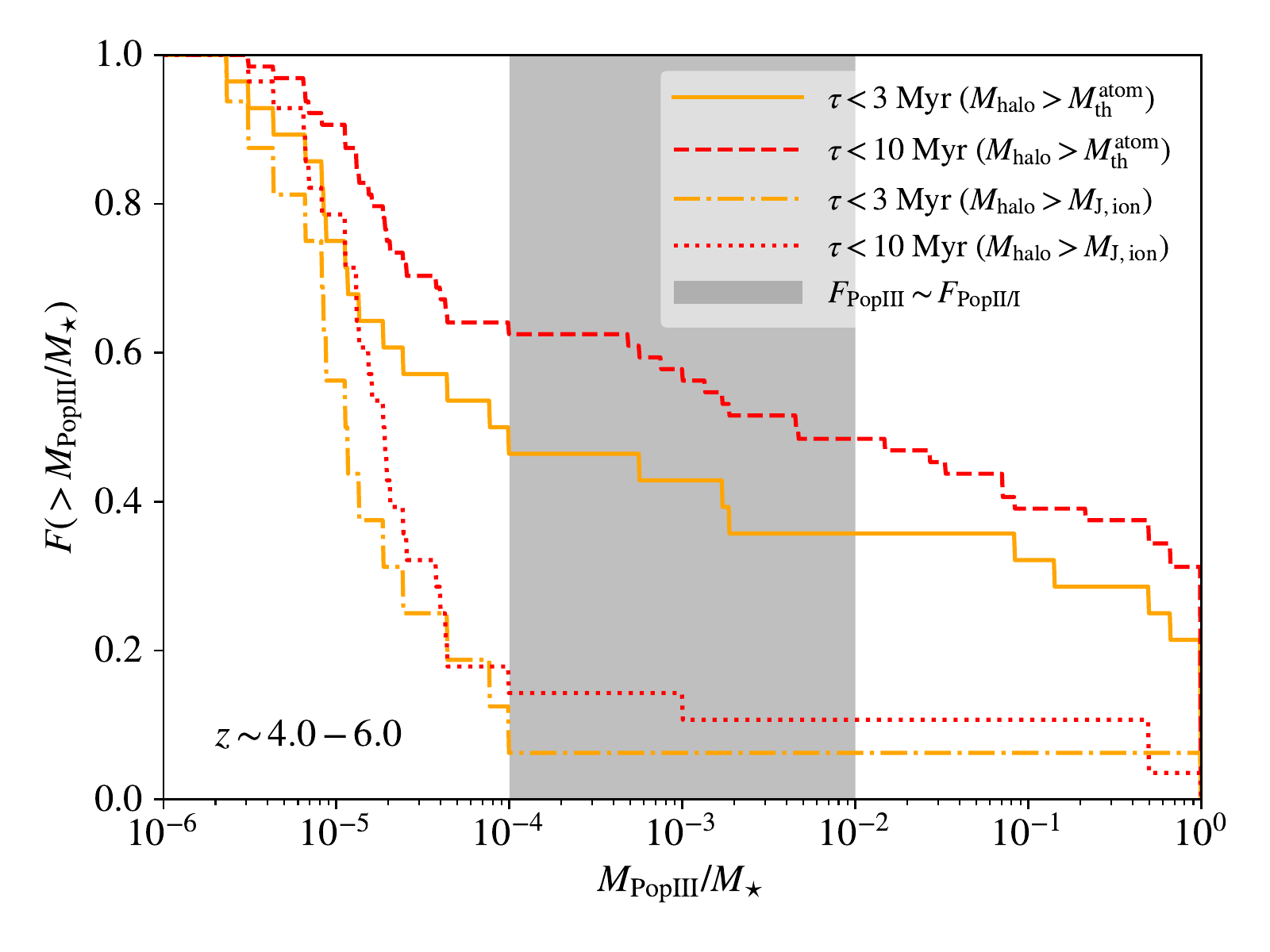}}
\vspace{-10pt}
\caption{Probability of detecting Pop~III-bright systems. Left panel (a): Fractions of star-forming galaxies hosting active Pop~III stars. The results for all atomic cooling haloes are shown with the solid and dashed curves, while those for haloes above the Jeans mass of ionized gas with triangles and filled circles, for Pop~III ages $\tau<3$ and 10~Myr, respectively. 
Right Panel (b): Cumulative distribution functions of the active Pop~III to total stellar mass ratio, for atomic cooling haloes (solid and dashed) and haloes above the ionization Jeans mass (dashed-dotted and dotted) from the representative sample. The minimum ratio $M_{\rm PopIII}/M_{\star}\sim 10^{-4}-0.01$ for Pop~III stars to dominate the flux (i.e. $F_{\rm PopIII}\gtrsim F_{\rm PopII/I}$) is shown with the shaded region. }
\label{focc}
\end{figure*}

We can also consider the termination of Pop~III star formation from the observational perspective. Given the host properties and Pop~III SFRD models in Sec.~\ref{s4.1} and \ref{s4.2}, we can now predict the detection rates of Pop~III stars and their PISNe as functions of the horizon redshift $z_{\rm PopIII}$, as shown in Fig.~\ref{Drate}. Here, we adopt a field of view (FoV, i.e. survey area) of $10\ \rm arcmin^{2}$ for direct observation of Pop~III stars, which is relevant to JWST and Hubble deep-field campaigns, and $\rm FoV=10\ \rm deg^{2}$ for detection of Pop~III PISNe, achievable with the \textit{Vera C. Rubin Observatory}, specifically its \textit{Legacy Survey of Space and Time} (LSST). 

In the optimistic case, where metal mixing is inefficient ($\beta_{\rm mix}\sim 0$, $\langle f_{\rm mp}\rangle\sim 1$), direct detection of Pop~III systems would reach $\sim 10\ \rm arcmin^{-2}$ for $z_{\rm PopIII}\sim 2$, and up to 2000 per $\rm arcmin^{2}$ for $z_{\rm PopIII}\sim 10$, assuming that all Pop~III stars are grouped into systems of $M_{\rm PopIII}= 10^{3}\ \rm M_{\odot}$. However, as the simulated Pop~III systems are not massive ($M_{\rm PopIII}\lesssim 2\times 10^{3}\ \rm M_{\odot}$), we infer $z_{\rm PopIII}\sim 0.5$ for the \textit{Hubble Space Telescope} (HST) and JWST, leading to a detection rate $\sim 0.1$ per $\rm arcmin^{2}$ even in the optimistic case. Here, in the calculation of $z_{\rm PopIII}$, we consider the HST WFC3 filter F555W with a limiting (AB) magnitude of 30 (for $\rm SNR>5$ in a 10-hour exposure), and the JWST NIRCam filter F150W with a limiting magnitude of 31.4. Optimistic magnitudes for Pop~III stellar systems are derived with the Stellar Population Synthesis (SPS) code \textsc{yggdrasil}\footnote{\url{https://www.astro.uu.se/~ez/yggdrasil/yggdrasil.html}} \citep{zackrisson2011spectral}, under their (instantaneous-burst) Pop~III.1 model (with an extremely top-heavy Salpeter IMF in the range of $50-500\ \rm M_{\odot}$) based on \citet{schaerer2002properties}, and optimal parameters for nebular emission and Ly$\alpha$ transmission (i.e. $f_{\rm cov}=1$, $f_{\rm Ly\alpha}=0.5$). 

Using their Pop~III.2 model with a moderately top-heavy IMF from \citet{raiter2010predicted} will reduce the flux by a factor of 3. If lensing pushes the limiting magnitude to 33, we can reach $z_{\rm PopIII}\sim 4$ with JWST, where up dozens of Pop~III host systems will reside in one $\rm arcmin^{2}$, but the fraction of lensed systems may still be too low for promising detection. For the \textit{Wide Field Infrared Survey Telescope} (WFIRST), given a sensitivity similar to that of HST in the optical and a large FoV of $0.3\ \rm deg^{2}\approx 10^{3}\ arcmin^{2}$, one exposure of 10 hours can detect $\sim 100$ Pop~III systems with $\rm SNR>5$ (for the R062 filter) at $z\lesssim 0.5$ in the optimistic case. However, in the pessmimistic model with halo-scale metal mixing of \textit{simulated} haloes (i.e. $\langle f_{\rm mp}\rangle=\hat{f}_{\rm mp}^{\rm sim}$), the detection rate will be reduced by a factor of 30. Furthermore, as discussed below, it is non-trivial to identify Pop~III host systems at such low redshifts when Pop~II/I star formation is dominating and prevalent. This is particularly challenging at low redshifts ($z\lesssim 1$), where no powerful instrument currently exists in the rest-frame UV to search for distinct features of Pop~III (e.g. bluer spectra and \HeII\ emission lines).

Beside direct observation of active Pop~III stars themselves, detection of Pop~III PISNe is another important channel to probe late-time Pop~III star formation \citep{scannapieco2005detectability}. However, we find that the scarcity of Pop~III stars remains the main obstacle to detection of their PISNe, as seen in previous studies (e.g. \citet{hummel2012}). Even for a FoV as large as $10\ \rm deg^{2}$, the detection rate only reaches 1 per year at $z_{\rm PopIII}\sim 7$ in the optimistic model. While the estimated horizon redshift is $z_{\rm PopIII}\sim 0.75-2$ for LSST\footnote{The range of $z_{\rm PopIII}$ for LSST reflects the uncertainty in how PISN blast waves interact with the circumstellar medium \citep{nicholl2020}.}, and that for JWST NIRSpec ($\rm SNR>10$) is $z_{\rm PopIII}\sim 5$, based on the properties of the recently discovered PISN candidate SN2016aps \citep{nicholl2020}. For LSST, in the optimistic case with $z_{\rm PopIII}=2$, where all Pop~III PISNe are massive interacting events similar to SN2016aps \citep{nicholl2020} and luminous for long ($\sim 1$~yr), a survey area of $\gtrsim 100\ \rm deg^{2}$ is required to detect one Pop~III PISN. For JWST, although it can reach $z_{\rm PopIII}\sim 5$, it cannot afford a large survey area. For instance, the FoV considered by the First Lights at REionization (FLARE) project is only $\sim 0.1\ \rm deg^{2}$, such that detection of Pop~III PISNe is unlikely to be achieved in a survey time of a few years \citep{wang2017first,regHos2020detecting}. In our calculation of the PISN rates, we assume a Pop~III PISN efficiency $\epsilon_{\rm PISN}=10^{-3}\ \rm M_{\odot}^{-1}$ for typical top-heavy Pop~III IMFs. We adopt $\epsilon_{\rm PISN}=6\times 10^{-5}\ \rm M_{\odot}^{-1}$ for Pop~II/I as an optimal estimation, based on a Salpeter IMF from 0.1 to 200 $\rm M_{\odot}$, neglecting mass loss from stellar winds. Given such assumptions and the Pop~II/I SFRD from \citet{madau2014araa}, we find that the Pop~III contribution to the total PISN rate remains below $10^{-3}$ at $z\lesssim 7$. 
Tuning the Pop~III IMF and normalization of the Pop~III SFRD within observational constraints (e.g. \citealt{visbal2015,inayoshi2016gravitational}) can enhance the Pop~III PISN rate by at most a factor of 30, such that the dominance of Pop~II/I remains, unless the (average) PISN efficiency for Pop~II/I stars is much lower in reality than our optimal estimation. Note that the PISN efficiency is highly sensitive to the upper mass limit of a stellar population, which is still uncertain, especially for Pop~III and II stars with low metallicities. For metal-enriched stars (Pop~II/I), strong stellar winds may drive the upper mass limit below the PISN threshold. In that case, {\it only} Pop~III would contribute to the PISN rate. 

In general, our results indicate that detection of Pop~III stars and PISNe in the post-reionization epoch is extremely challenging, even for the optimistic model with continuous Pop~III star formation at a rate $\sim 10^{-7}-10^{-4}\ \rm M_{\odot}\ yr^{-1}\ Mpc^{-3}$ down to $z\sim 0$. 
Actually, this prospect will be rendered even more difficult, if we further consider the fact that most Pop~III stars formed at late times ($z\lesssim 6$) would reside in massive systems, where (young massive) Pop~II/I stars are also present. As long as $M_{\rm PopIII}\lesssim 10^{3}\ \rm M_{\odot}$, any galaxy with a Pop~II/I SFR $\dot{M}_{\star}\gtrsim 10^{-2}\ \rm M_{\odot}\ yr^{-1}$ in the past 10 Myr or a total stellar mass $M_{\star}\gtrsim 10^{7}\ \rm M_{\odot}$ will be dominated by the light from Pop~II/I stars, even if it has experienced recent Pop~III star formation. In other words, the active Pop~III to total stellar mass ratio $M_{\rm PopIII}/M_{\star}$ must be above $\sim 10^{-4}-0.01$ for the Pop~III to Pop~II/I flux ratio to exceed one ($F_{\rm PopIII}\gtrsim F_{\rm PopII/I}$). We regard such systems as Pop~III-bright\footnote{We also use \textsc{Yggdrasil} to derive the magnitudes of Pop~II/I stars, with a Kroupa IMF in the range of $0.1-100\ \rm M_{\odot}$, a metallicity $Z=0.02\ \rm Z_{\odot}$ and a constant SFR over 10~Myr, based on the Starburst99 Padova-AGB tracks \citep{leitherer1999starburst99,vazquez2005optimization}. Again, optimal parameters for nebular emission and Ly$\alpha$ transmission are adopted (i.e. $f_{\rm cov}=1$, $f_{\rm Ly\alpha}=0.5$).}.

In Fig.~\ref{focc}, we explore the probability of identifying Pop~III-bright systems in dwarf galaxies ($M_{\star}\lesssim 10^{8.5}\ \rm M_{\odot}$) by considering the occupation fraction of Pop~III hosts in star forming galaxies (left), and the cumulative distribution function of the active Pop~III to total stellar mass ratio (right), for the representative sample. Before reionization ($z\gtrsim 10$), $\sim 5$ (10)\% of all atomic cooling haloes have recent Pop~III activities within 3 (10) Myr. However, after reionization ($z\sim 4-6$), only $\sim 1$ (2)\% percent of star-forming haloes with $M_{\rm halo}>M_{\rm J,ion}$ host active Pop~III stars for $\tau<3$ (10)~Myr, and the occupation fraction will decrease with decreasing redshift, similar to the trend in Pop~III SFRD. Moreover, given the Pop~III-bright criterion $M_{\rm PopIII}/M_{\star}\gtrsim 10^{-4}-0.01$, $35-60$\% of the atomic cooling haloes with recent Pop~III star formation are Pop~III-bright, while only $\lesssim 10$\% of Pop~III host haloes with $M_{\rm halo}>M_{\rm J,ion}$ are Pop~III-bright\footnote{Note that our simulation is limited in volume such that only dwarf galaxies ($M_{\star}\lesssim 10^{8.5}\ \rm M_{\odot}$) at $z\sim 4-6$ are considered in this analysis. The fraction of Pop~III-bright systems is expected to be lower at lower redshifts, where more massive galaxies will be the (potential) hosts of Pop~III stars. Such massive galaxies are more likely have dominant Pop~II/I components, where feedback from the central massive black holes can also regulate Pop~III star formation.}. As a result, Pop~II/I stars will dominate in most ($\gtrsim 99.9\%$) massive haloes ($M_{\rm halo}\gtrsim M_{\rm J,ion}\sim 10^{9}\ \rm M_{\odot}$) after reionization ($z\lesssim 6$), according to the star formation main sequence and assembly histories of such haloes \citep{pawlik2013first,sparre2015star,yajima2017growth}, whereas before reionization ($z\gtrsim 10$), $\sim 2.5-6$\% of all dwarf galaxies in atomic cooling haloes will be Pop~III-bright.

However, given $M_{\rm PopIII}< 10^{4}\ \rm M_{\odot}$, such galaxies must form less than $10^{6}\ \rm M_{\odot}$ Pop~II/I stars within 10 Myr, such that they cannot be reached by JWST at $z\gtrsim 6.5$. 
Nevertheless, as mentioned in Sec.~\ref{s4.2}, the total mass of active Pop~III stars per halo/galaxy itself is still uncertain, which depends on resolution and the sub-grid models for star formation and stellar feedback, particularly chemical feedback from SNe. In the optimal case where $M_{\rm PopIII}\sim 10^{5}\ \rm M_{\odot}$ 
, Pop~III-bright galaxies would be detectable by JWST (HST/WFIRST) up to $z\sim 12\ (4)$. We thus arrive at the conclusion that right before reionization ($z\sim 10$) is the optimal epoch to search for Pop~III-bright systems, consistent with \citet{sarmento2018following}, which predict a Pop~III-bright\footnote{Note that \citet{sarmento2018following} adopts a more strict definition for `Pop~III-bright' as $F_{\rm PopIII}>3F_{\rm PopII/I}$. Therefore, the values quoted here should be regarded as lower limits for our definition ($F_{\rm PopIII}>F_{\rm PopII/I}$). } fraction of $\sim 2.5-16$\% at $z\sim 9-10$. In this way, our optimistic Pop~III SFRD model predicts that JWST (HST/WFIRST) is able to find $\sim 10$~(0.1) such Pop~III-bright systems per $\rm arcmin^{2}$. Again, in more realistic models with enhanced metal mixing, those detection rates would be significantly suppressed.

\section{Summary and Conclusions}
\label{s5}
We construct a theoretical framework to study Pop~III star formation in the post-reionization epoch ($z\lesssim 6$) by combining cosmological simulation data with semi-analytical models. To be specific, we closely look into a representative sample of haloes hosting active Pop~III stars at $z\sim 4-6$ from a cosmological simulation in LB20 (Sec.~\ref{s2} and \ref{s4.1}). Based on this, we extrapolate the Pop~III SFRD to $z=0$ with additional semi-analytical modelling of turbulent metal mixing and reionization (Sec.~\ref{s4.2}), which may not be fully captured in the simulation. In this way, we evaluate the key physical processes that shape Pop~III star formation at late times and the corresponding observational prospects (Sec.~\ref{s4.3}). Although many of these processes are currently not well understood, future theoretical and observational efforts will reduce the uncertainties and shed light on the fundamental question of the termination of Pop~III star formation. Our main findings are summarized below.
\begin{itemize}
    \item Both radiative and chemical feedback play important roles in regulating Pop~III star formation. The former, in terms of LW feedback and reionization, shifts (potential) Pop~III star formation to massive haloes (i.e. atomic cooling haloes and haloes above the filtering mass, $M_{\rm halo}\gtrsim 10^{7-9}\ \rm M_{\odot}$). The latter, in terms of mixing of metals released from SNe into the interstellar/circumgalactic medium (ISM/CGM), then determines whether Pop~III star formation is possible or not in such massive haloes, which is particularly important in the post-reionization epoch. 
    
        In our optimistic model (without additional metal mixing beyond that captured by the simulation), the contribution of minihaloes (i.e. the `classical' site of Pop~III star formation) to the overall Pop~III SFRD drops to a few percent at $z\lesssim 13$ due to LW feedback (see Fig.~\ref{fiso}), and decreases exponentially with redshift (to $\lesssim 10^{-5}$ at $z=0$) after reionization ($z\lesssim 6$). Late-time Pop~III star formation is dominated by massive haloes ($M_{\rm halo}\gtrsim 10^{9}\ \rm M_{\odot}$), where the densest regions have been significantly metal enriched, but pockets of dense metal-poor gas (e.g. in cold accretion flows) may still exist to form Pop~III stars due to inefficient metal mixing (see Fig.~\ref{gasdis}), consistent with the `Pop~III wave' theory \citep{tornatore2007population}. 
        
        However, limited by resolution, treatments of metal mixing are imperfect in cosmological simulations, such that metal mixing can be more efficient in reality than in our optimistic model (see Sec.~\ref{s3.2} for details). For instance, if we assume that metals are fully mixed within the halo virial radius, the Pop~III SFRD would be reduced by more than a factor of 30 at $z\lesssim 6$. 
    
    \item Next to the global Pop~III SFRD, the metal mixing process is also important for another key parameter, the total mass, $M_{\rm PopIII}$, of \textit{active} Pop III stars per host halo. 
    Note that we here focus on \textit{active} Pop~III stars, and the relevant timescale is short (a few Myr) by nature, such that the signals of short-lived massive Pop~III stars can add up in observation. Therefore, $M_{\rm PopIII}$ is equivalent to the (instantaneous) Pop~III SFR measured at a timescale of a few Myr. 
    In general, $M_{\rm PopIII}$ is the product of the `quantum' of Pop~III star formation, i.e. the typical Pop~III stellar mass formed per local (cloud-scale) star formation event, and the number of dense ($n_{\rm H}\gtrsim 100\ \rm cm^{-3}$) metal-poor ($Z\lesssim 10^{-6}-10^{-3.5}\ \rm Z_{\odot}$) star-forming gas clouds in the ISM/CGM of a halo, coexisting on a timescale of a few Myr. 
    The former is well constrained to $500-1000\ \rm M_{\odot}$ for $\Lambda$CDM\footnote{The picture can be different for other dark matter models (see e.g. \citealt{gao2007lighting,hirano2017first}).} by high-resolution simulations and constraints from the timing of the global 21-cm absorption signal \citep{stacy2013constraining,susa2014mass,machida2015accretion,stacy2016building,hirano2017formation,Anna2018,hosokawa2020}. While the latter is highly sensitive to the metal mixing process. 
    

        Therefore, in $\Lambda$CDM, $M_{\rm PopIII}$ reflects the number of newly-formed Pop~III star clusters in a halo. 
        In our simulation, we find that only a few Pop~III clusters can be formed within a few Myr per halo, i.e. $M_{\rm PopIII}<10^{4}\ \rm M_{\odot}$ and the average is $\langle M_{\rm PopIII}\rangle\simeq 10^{3}\ \rm M_{\odot}$ (see Fig.~\ref{MpopIII}). Interestingly, we also find that $M_{\rm PopIII}$ is independent of halo mass and total stellar mass, quite different from the case of Pop~II/I stars where SFR is correlated with stellar mass (i.e. the star formation main sequence). This further indicates the importance of metal mixing for Pop~III star formation.
        
    \item The total mass $M_{\rm PopIII}$ is particularly important for direct detection of Pop~III stars. For instance, if $M_{\rm PopIII}\sim 10^{3}\ \rm M_{\odot}$, as shown in our simulation, direct detection of Pop~III stars is only possible at very low redshifts ($z\lesssim 0.5$), considering the sensitivities of space telescopes at present or in the near future (e.g. HST, JWST and WFIRST). If Pop~III star formation were to extend to such low redshifts, as predicted by our optimistic model, WFIRST, with its large FoV, could detect $\sim 100$ galaxies with active Pop~III stars in one exposure of 10 hours. However, as long as $M_{\rm PopIII}\sim 10^{3}\ \rm M_{\odot}$, only the faintest hosts of Pop~III stars can be identified as Pop~III-bright (where the Pop~III flux exceeds that of Pop~II/I), while the emission from the majority ($\gtrsim 99.9$\%) of luminous hosts will be dominated by Pop~II/I stars, unless we observe in the rest-frame UV. Unfortunately, no powerful UV instrument currently exists to search for distinct features of Pop~III (e.g. bluer spectra and \HeII\ emission lines) in the rest-frame UV at such low redshifts. Detection of Pop~III-bright systems would still be challenging for WFIRST.
        
        Nevertheless, as metal mixing is not well understood, $M_{\rm PopIII}$ is still uncertain (see \citealt{xu2016late,danielle2020,pallottini2014simulating,sarmento2018following}). Our value lies at the lower end, while the upper limit is $\sim 10^{6}\ \rm M_{\odot}$, derived from theoretical calculations of collapsing primordial gas \citep{yajima2017upper} and the recent non-detection of Pop~III features in the Hubble Frontier Fields at $z\sim 6-9$ \citep{bhatawdekar2020}. If metal-mixing is overestimated in our simulation and $M_{\rm PopIII}\sim 10^{5-6}\ \rm M_{\odot}$ in reality, Pop~III-bright galaxies will be detectable by JWST (HST/WFIRST) up to $z\sim 12.5\ (5)$. In this way, our optimistic Pop~III SFRD model predicts that JWST (HST/WFIRST) is able to find up to $\sim 10$~(0.1) Pop~III-bright systems per $\rm arcmin^{2}$. 
    
    \item Finally, our simulations, similar to previous cosmological simulations \citep{wise2011birth,johnson2013first,pallottini2014simulating,xu2016late}, predicts that the overall volume-filling fraction of metal-enriched gas is only a few percent when the universe has expanded to 10-20\% of its current size. As it is more difficult to enrich large volumes of gas when the Universe expands further, such simulation results imply that the majority ($\gtrsim 90$\%) of the IGM in the observable Universe is occupied by metal-free gas, likely at very low column-densities, undetectable by current instruments. Detecting and quantifying this metal-free phase of the IGM will constrain theoretical models of metal mixing and, therefore, late-time Pop~III star formation. This is complemented by similar observations of metal-absorption lines at high redshifts, where bright gamma-ray burst afterglows could serve as background sources \citep{wang2012}. Our semi-analytical modelling for gravity-driven turbulent metal mixing in virialized systems can be easily extended to describe IGM metal enrichment, governed by the same mixing strength parameter, $\beta_{\rm mix}$ (and additional parameters if necessary). Therefore, it is possible to directly relate late-time Pop~III star formation to the volume-filling fraction of metal-free gas. We defer such exploration to future work.
\end{itemize}

When did Pop~III star formation end? The current answer is uncertain. In the optimistic case, Pop~III star formation would extend to $z\sim 0$ at a low yet non-negligible rate of $\sim 10^{-7}\ \rm M_{\odot}\ yr^{-1}\ Mpc^{-3}$, while in the pessimistic case, Pop~III star formation may already be terminated by the end of reionization ($z\sim 5$). 
To answer this fundamental question, we must better understand cosmic chemical evolution in terms of mixing of metals released by SNe into the ISM/CGM/IGM during structure formation. On the theory side, we need cosmological simulations with proper resolution and complete representations of the halo population (from minihaloes $\sim 10^{6}\ \rm M_{\odot}$ to galaxy clusters $\sim 10^{14}\ \rm M_{\odot}$) across the entire cosmic history, equipped with advanced sub-grid models for metal mixing and zoom-in simulation techniques (e.g. \citealt{pan2013modeling,hopkins2017anisotropic,stephen2020}). For observations, stronger constraints will soon come from JWST and WFIRST for (potential) high-$z$ and low-$z$ sources, on both the overall Pop~III SFRD and the typical total mass of active Pop~III stars per halo. Gravitational wave observations of the binary black hole mergers originated from Pop~III stars can also constrain the Pop~III SFRD (e.g. \citealt{sesana2009observing,kinugawa2014possible,hartwig2016,belczynski2017likelihood,liu2020}). Meanwhile, we advocate for new UV space telescopes to search for galaxies with distinct Pop~III features in the rest-frame UV at low redshifts ($z\lesssim 1$), and programs designed to measure the volume-filling fraction of metal-free gas in the IGM. 

Overall, many aspects regarding Pop~III star formation are still uncertain, as discussed here via our framework. Nevertheless, with improved theoretical and observational efforts, particularly on the metal mixing process\footnote{Metal mixing is also crucial for inferring the properties of Pop~III stars from observations of extremely metal-poor stars in the local Universe, i.e. `stellar archaeology' (e.g. \citealt{frebel2015near,ji2015preserving,hartwig2015,ishigaki2018initial,magg2019observational,magg2020}).}, we will arrive at a more complete picture of Pop~III star formation, from onset to termination, thus further elucidating the most elusive population of stars. 

\section*{Acknowledgements}
The authors wish to thank Mar\'ia Emilia De Rossi for insightful discussion regarding stellar population synthesis models for Pop~III stars, and acknowledge the Texas Advanced Computing Center (TACC) for providing HPC resources under XSEDE allocation TG-AST120024. 

\section*{Data availability}
The data underlying this article will be shared on reasonable request to the corresponding authors.

\bibliographystyle{mnras}
\bibliography{ref} 


\bsp	
\label{lastpage}

\end{document}